%% file: main.tex
\documentclass[final,5p,times]{elsarticle}

\journal{Future Generation Computer Systems}
\input{styles.tex}

\input{macro.tex}

\newcommand{\numusecases}{21\xspace}

\newtoggle{draft}
\togglefalse{draft}

\makeatletter
\renewcommand\paragraph{\@startsection{paragraph}{4}{\z@}%
  {2.25ex \@plus 1ex \@minus .2ex}%
  {-0.75em}%
  {\normalfont\normalsize\bfseries}}
\makeatother

\makeatletter
\lstdefinestyle{mystyle}{
  basicstyle=%
    \ttfamily
    \lst@ifdisplaystyle\scriptsize\fi
}
\makeatother

\begin{document}
\hyphenation{block-chain}
\hyphenation{Block-chain}
\hyphenation{block-chains}
\begin{frontmatter}

\title{Smart Contract Languages: a comparative analysis}

\author[1]{Massimo Bartoletti\corref{bart}}
\author[2]{Lorenzo Benetollo}
\author[2]{Michele Bugliesi}
\author[4]{Silvia Crafa}
\author[4]{Giacomo Dal Sasso}
\author[3]{Roberto Pettinau}
\author[1]{Andrea Pinna}
\author[1]{Mattia Piras}
\author[2]{Sabina Rossi}
\author[1]{Stefano Salis}
\author[2]{Alvise Span\`o}
\author[1]{Viacheslav Tkachenko}
\author[1]{Roberto Tonelli}
\author[5]{Roberto Zunino}

\cortext[bart]{\emph{Corresponding author}. Dipartimento di Matematica e Informatica, Universit\`a degli Studi di Cagliari, via Ospedale 72, 09124 Cagliari (Italy), e-mail: \texttt{bart@unica.it}}

\affiliation[1]{organization={Universit\`a degli Studi di Cagliari},
            city={Cagliari},
            country={Italy}}

\affiliation[2]{organization={Universit\`a Ca' Foscari Venezia},
            city={Venezia},
            country={Italy}}

\affiliation[3]{organization={Technical University of Denmark},
            city={Copenhagen},
            country={Denmark}}

\affiliation[4]{organization={Universit\`a di Padova},
            city={Padova},
            country={Italy}}

\affiliation[5]{organization={Universit\`a degli Studi di Trento},
            city={Trento},
            country={Italy}}

\input{abstract.tex}


\end{frontmatter}

\input{intro.tex}

\input{overview.tex}

\input{languages.tex}

\input{solidity-ethereum.tex}

\input{rust-solana.tex}
\input{aiken-cardano.tex}

\input{pyteal-algorand.tex}

\input{move-aptos.tex}

\input{smartpy-tezos.tex}
\input{comparison.tex}

\input{tab-strengths.tex}
\input{related.tex}
\input{conclusions.tex}

\paragraph{Acknowledgments}
Work partially supported by the Project PRIN 2020  “Nirvana - Noninterference and Reversibility Analysis in Private Blockchains” and by projects PRIN 2022 DeLiCE (F53D23009130001)  and SERICS (PE00000014) under the MUR National Recovery and Resilience Plan funded by the European Union - NextGenerationEU.
The authors declare that they had no investment or advisory relationships with any
of the blockchain companies/foundations cited in this research.

\bibliography{main} 

\end{document}

%% file: styles.tex
\usepackage[T1]{fontenc}

\usepackage{lipsum}

\usepackage{color}
\usepackage[usenames,dvipsnames]{xcolor}

\usepackage{fontawesome} 

\usepackage{multicol}
\usepackage{booktabs}
\usepackage{siunitx}
\usepackage{textcomp}
\usepackage[labelfont=bf]{caption}
\usepackage{amssymb}
\usepackage{graphicx}
\usepackage{svg}

\usepackage{paralist}
\usepackage{environ}
\usepackage{tikz}
\usetikzlibrary{shadows}
\usetikzlibrary{matrix,chains,positioning,decorations.pathreplacing}
\usetikzlibrary{patterns,calc,fit,arrows}
\usetikzlibrary{shapes.geometric}

\usepackage{mathtools}
\usepackage{array,multirow}
\usepackage{adjustbox}
\usepackage{algorithm}
\usepackage[hyphens]{url}

\usepackage{mdframed}
\usepackage{arydshln}
\usepackage{etoolbox}
\usepackage{xspace} 
\usepackage{dirtytalk} 
 
\usepackage{makecell}
\usepackage{stackengine}
\setstackEOL{\cr} 

\usepackage[inline,shortlabels]{enumitem} 
\newlist{inlinelist}{enumerate*}{1}
\setlist*[inlinelist,1]{%
  label=(\roman*),
}

\usepackage{caption}
\usepackage{lstautogobble}
\usepackage{rotating}

\usepackage{nicefrac}

\usepackage[normalem]{ulem} 

\usepackage{xifthen}  
\usepackage{ifthen}

\usepackage[hidelinks]{hyperref}
\usepackage{cleveref}
\usepackage{bigints}
\usepackage{showkeys}
\usepackage[final,commandnameprefix=always,defaultcolor=red]{changes}

\usepackage[final,nomargin,inline,index]{fixme} 
\fxusetheme{color}
\definecolor{alvicolor}{rgb}{0.45, 0.20, 0.90}
\FXRegisterAuthor{bart}{anbart}{\color{magenta} {\underline{bart}}}
\FXRegisterAuthor{andrea}{anand}{\color{red} {\underline{andrea}}}
\FXRegisterAuthor{tonelli}{antonelli}{\color{blue} {\underline{tonelli}}}
\FXRegisterAuthor{alvi}{analvi}{\color{alvicolor} {\underline{alvi}}}
\FXRegisterAuthor{lore}{anbntl}{\color{orange} {\underline{lore}}}
\FXRegisterAuthor{silvia}{ansil}{\color{purple} {\underline{silvia}}}
\FXRegisterAuthor{viach}{anviach}{\color{teal} {\underline{viach}}}
\FXRegisterAuthor{giaco}{angiaco}{\color{gray} {\underline{giaco}}}
\FXRegisterAuthor{robpe}{anrobpe}{\color{teal} {\underline{robpe}}}
\FXRegisterAuthor{stefano}{anstefano}{\color{cyan} {\underline{stefano}}}
\FXRegisterAuthor{mik}{anmik}{\color{violet} {\underline{mik}}}
\FXRegisterAuthor{sabina}{ansab}{\color{ForestGreen} {\underline{sabina}}}
\FXRegisterAuthor{zun}{anzun}{\color{MidnightBlue} {\underline{zun}}}

\hypersetup{
  breaklinks   = true,
  colorlinks   = true, 
  urlcolor     = blue, 
  linkcolor    = blue, 
  citecolor    = red   
}
\hypersetup{final} 

\usepackage{listings}
\definecolor{listingBG}{HTML}{FFFFCB}%
\definecolor{listingFrame}{HTML}{BBBB98}%
\definecolor{listingLineno}{rgb}{0.5,0.5,1.0}%

\definecolor{LightGrey}{rgb}{0.975,0.975,0.975}

\lstdefinelanguage{solidity}{
	commentstyle=\color{Gray},
	morecomment=[l]{//},
	morecomment=[s]{/*}{*/},
	classoffset=0,
        escapechar=\$,
	morekeywords={struct,mapping,function,this,public,private,static,final,class,extends,switch,case,break,finally,try,catch,return,if,else,new},
	keywordstyle=\color{NavyBlue}\bfseries,
	classoffset=1,
	morekeywords={unit,int,string,bool,address,uint,uint256},
	keywordstyle=\color{blue},
	classoffset=2,
	morekeywords={ether,wei,finney,contract,send,throw,msg,sender,value},
	keywordstyle=\color{Plum}\bfseries,
}

\lstdefinelanguage{move}{
	commentstyle=\color{Gray},
	morecomment=[l]{//},
	morecomment=[s]{/*}{*/},
	classoffset=0,
        escapechar=\$,
	morekeywords={let,struct,public,if,else,new,fun,acquires,phantom,has,assert,mut},
	keywordstyle=\color{NavyBlue}\bfseries,  
	classoffset=1,
	morekeywords={u64,u8,bool,address,key,store,copy,drop,signer,Coin},
	keywordstyle=\color{blue},
	classoffset=2,
	morekeywords={move_to,move_from,borrow_global,borrow_global_mut,exists,merge,deposit,withdraw,address_of},
	keywordstyle=\color{Plum}\bfseries,
}

\def\addrColor{\color{magenta}}

\lstdefinelanguage{pseudocode}{
	commentstyle=\color{Gray},
	morecomment=[l]{//},
	morecomment=[s]{/*}{*/},
	classoffset=0,
        escapeinside={(*}{*)},
	morekeywords={if,then,else,elif,contract,skip,require,abort,return,send,expect,owns,isSigner},
	keywordstyle=\color{Plum},
	classoffset=1,
	morekeywords={var,T}, 
	keywordstyle=\addrColor,
	classoffset=2,        
	morekeywords={},
	keywordstyle=\pmvColor,
	classoffset=3,        
        morekeywords={constructor},
	keywordstyle=\color{MidnightBlue}\bfseries,
	basicstyle=\fontseries{m}\normalsize\ttfamily
	\lst@ifdisplaystyle\footnotesize\fi,
}

\lstdefinelanguage{rust}{
    commentstyle=\color{Gray},
    morecomment=[l]{//},
    morecomment=[s]{/*}{*/},
    classoffset=0,
    escapechar=\$,
    morekeywords={fn, let, mut, match, if, else, use, pub, impl, struct, enum, trait, self, true, false, require},
    keywordstyle=\color{NavyBlue}\bfseries,
    classoffset=1,
    morekeywords={u64, i32, bool, String, Vec, Option, Result, Some, None, Ok, Err},
    keywordstyle=\color{blue},
    classoffset=2,
    morekeywords={self, crate, super, mod, as, break, continue, return, loop},
    keywordstyle=\color{Plum}\bfseries,
    sensitive=true,
    morestring=[b]"
}

%% file: macro.tex
\newcommand{\ifempty}[3]{%
  \ifthenelse{\isempty{#1}}{#2}{#3}%
}

\def\negcaptionspace{\vspace{-10pt}}

\usepackage{refcount}

\newcommand{\myfootnotetext}[1]{\footnotetext{#1\label{fn:text}%
        \edef\fnmark{\getpagerefnumber{fn:mark}}%
        \edef\fntext{\getpagerefnumber{fn:text}}%
        \ifx\fnmark\fntext\else\ClassWarning{}{footnote mark and text on different pages!}\fi}}

\newcommand{\codefont}{\fontsize{10}{10}\selectfont}
\newcommand{\code}[1]{{\tt\codefont {#1}}}
\newcommand{\smallcode}[1]{{\tt\codefont\footnotesize {#1}}}
\newcommand{\contract}[1]{{\tt\codefont{\txColor{#1}}}}
\newcommand{\txcode}[1]{{\text{\tt\codefont{\txColor{#1}}}}}
\newcommand{\smalltxcode}[1]{{\text{\tt\codefont\footnotesize{\txColor{#1}}}}}
\newcommand{\solkeyword}[1]{{\text{\tt\codefont{\color{Plum}{#1}}}}}

\def\etc{\emph{etc}.\@\xspace}
\newcommand{\Eg}{E.g.\@\xspace}
\newcommand{\eg}{e.g.\@\xspace}
\newcommand{\ie}{i.e.\@\xspace}
\newcommand{\wrt}{w.r.t.\@\xspace}

\newcommand{\keyterm}[1]{\textbf{\emph{#1}}}%

\newcommand{\tabrotdegrees}{90}
\newcommand{\tabrothspace}{0pt}

\newmdenv [linewidth=0pt]{mdNoFramed}
\def\txColor{\color{MidnightBlue}}
\def\fieldColor{\color{Plum}}
\newcommand{\txFmt}[1]{{\txColor{\sf #1}}}

\newcommand{\tx}[2][]{\txFmt{#2}_{\txColor{#1}}}
\newcommand{\txT}[1][]{\tx[#1]{T}} 

\def\pmvColor{\color{red}}
\newcommand{\pmvFmt}[1]{{\pmvColor{\tt #1}}}
\newcommand{\pmv}[2][]{\pmvFmt{#2}_{\pmvColor{#1}}\xspace}
\newcommand{\pmvA}[1][]{\pmv[{#1}]{A}} 
\newcommand{\pmvB}[1][]{\pmv[{#1}]{B}}
\newcommand{\pmvC}[1][]{\pmv[{#1}]{C}}

\DeclareMathSymbol{:}{\mathord}{operators}{"3A}
\def\tokColor{\color{magenta}}
\newcommand{\tokFmt}[1]{{ \mathtt{\tokColor{#1}} }}
\newcommand{\tok}[2][]{\tokFmt{#2}_{\tokColor{#1}}\xspace}
\newcommand{\tokT}[1][]{\tok[{#1}]{T}}
\newcommand{\tokTi}[1][]{\tok[{#1}]{T'}}

\def\fieldColor{\color{Plum}}
\newcommand{\txTag}[3][]{{\fieldColor\sf #3}\ifempty{#1}{\ifempty{#2}{}{: {#2}}}{[{#1}]\ifempty{#2}{}{: {#2}}}}

\newcommand{\txIn}[2][]{\txTag[{#1}]{#2}{in}}
\newcommand{\txWit}[2][]{\txTag[{#1}]{#2}{wit}}
\newcommand{\txOut}[2][]{\txTag[{#1}]{#2}{out}}
\newcommand{\txSigned}[2][]{\txTag[{#1}]{#2}{signed}}

\newcommand{\txscript}{\txTag{}{script}}
\newcommand{\txstorage}{\txTag{}{data}}
\newcommand{\txval}{\txTag{}{value}}

\newcommand{\rtx}{{\sf rtx}}

\newcommand{\native}{\checkmark}
\newcommand{\workaround}{\faCog\xspace}

%% file: abstract.tex
\begin{abstract}
\chadded{Smart contracts have played a pivotal role in the evolution of blockchains and Decentralized Applications (DApps). As DApps continue to gain widespread adoption, multiple smart contract languages have been and are being made available to developers, each with its distinctive features, strengths, and weaknesses. In this paper, we examine the smart contract languages used in major blockchain platforms, with the goal of providing a comprehensive assessment of their main properties. Our analysis targets the programming languages rather than the underlying architecture: as a result, while we do consider the interplay between language design and  blockchain model, our main focus remains on language-specific features such as usability, programming style, safety and security. To conduct our assessment, we propose an original benchmark which encompasses a wide, yet manageable, spectrum of key use cases that cut across all the smart contract languages under examination.}
\end{abstract}

%% file: intro.tex
\section{Introduction}

Smart contracts have played a pivotal role in the evolution of blockchain technology, paving the way for the emergence of the new paradigm of Decentralized Applications (DApps). As the DApps continue to gain popularity and become pervasive, the complexity of their business logic and the distributed, often open, nature of the underlying platforms over which they execute make their development an increasingly challenging task. 
In this article, we review the current advances in smart contract languages and assess them to gain fresh insights into their design principles and the impact on the programming practices they convey. 
\chadded{
Our analysis targets the programming languages rather than the underlying architectures, acknowledging that the design of robust smart contract languages is a prerequisite for a principled development of reliable and secure DApps.}

\paragraph{Methodology} 
\chadded{
 We start with an analysis of the tiered structure of blockchain platforms: our goal here is to single out the key architectural choices that affect the design and implementation of smart contracts. We then analyse and compare a selection of mainstream smart contract languages, based on an original benchmark we have developed to encompass a wide spectrum of key real-world DApp use cases~\cite{smart-contracts-comparison}.
}

To carry out our comparative analysis, we isolate six paradigmatic smart contract languages and their underlying blockchains --  Solidity on Ethereum, Rust on Solana, Aiken on Cardano, (Py)TEAL on Algorand, Move on Aptos, and SmartPy on Tezos -- as representatives of the permissionless platforms that have become mainstream and have gained widespread adoption in the development of DApps. While some of these platforms exist in different incarnations -- \eg Vyper is an alternative to Solidity on Ethereum, as Ligo is to SmartPy on Tezos and Plutus to Aiken on Cardano -- the results of our analysis remain largely consistent across these alternatives. In fact, languages operating on the same platform generally exhibit the same relevant properties relative to the features we target in our assessment, namely security, code readability, and usability. 

\chadded{The focus of our assessment is permissionless blockchains. Permissioned blockchains, in turn, are out of our present interests, as they usually come with general-purpose programming languages in which all the blockchain-specific features are managed within ad-hoc libraries that interact with the underlying blockchain consensus layer~\cite{Capocasale23bcra}.}

\paragraph{Main contributions}
Several analyses of blockchain platforms and smart contract languages have appeared in the recent literature (cf.~\Cref{sec:related}). One of the distinctive features of our present endeavor, one which sets it apart from previous experiments, is the hands-on nature of our experience with the use cases developed for benchmarking smart contracts.
The benchmark itself,  ``Rosetta Smart Contracts''~\cite{smart-contracts-comparison}, constitutes a major contribution, in that 
it encompasses a representative selection of common cases in DApp development that provides a smart contract \textit{chrestomathy}, the initial core of a standard test bed for a qualitative assessment of current and future smart contract languages (and platforms). 
Experimenting with the implementation of the use cases across the different languages proves very effective to enhance the understanding of the challenges in smart contract design, and of how such design is influenced by the tiered structure of the underlying blockchain. 
\chadded{
Specifically, we identify the choices made at the {\em contract layer} (as opposed to the lower, {\em consensus layer}) as the most influential for the design and the relevant properties of the overlying smart contract languages. At the contract layer, the blockchain is best understood as an asset-exchange state machine, where transactions activated by smart contract rules contribute to a state transition by either creating new assets or exchanging assets among users. Based on this view, we propose a categorization of smart contract languages based on the distinction between the two main models incarnating an asset-exchange state machine: the \emph{account-based model} and the \emph{UTXO model}. This perspective sheds new light on the interplay between the blockchain data and computational models on the one side, and the design principles of smart contracts on the other side. 
}

Our analysis also emphasizes the relevance of adequate language support for the key aspects of smart contract design: assets management, contract-to-contract interactions, and costs. Specifically, tailored type-level abstractions for creating, exchanging and operating with assets are a fundamental ingredient in preventing common errors and vulnerabilities such as asset loss, double spending, or unauthorized transfers. On a different, but related account, native support for certain functionalities of the underlying platform (\eg, custom tokens) is pivotal for key properties in security as well as in efficiency.

\paragraph{Structure of the paper}
We start in~\Cref{sec:overview} with an overview of smart contract platforms. Besides serving to set a common terminology for the analysis, this section also highlights how the basic choices at the contract layer influence smart contracts development, security and performance. We demonstrate this by discussing the (pseudo-code) implementation of a common use case in the account-based model (both in its stateful and stateless incarnations) and for the UTXO model. 
In~\Cref{sec:languages} we take a brief tour of the six smart contract languages in our selection, discussing their main features.
%
The core of the paper is~\Cref{sec:comparison}, where we develop our hands-on comparative 
analysis.
In~\Cref{sec:related} we contextualise our contribution in the scientific literature. Finally, in~\Cref{sec:conclusions} we conclude with a discussion of the key insights derived by our analysis.

%% file: overview.tex
\section{Smart contracts on blockchains}
\label{sec:overview}

Blockchain smart contracts are best understood as collections of executable rules that are triggered by user \keyterm{transactions} to activate the exchange of assets and other forms of interaction between users.
The underlying architecture is a tiered structure comprising two main layers\footnote{Blockchains are typically described as comprising more layers, including, from the bottom up, \emph{network, consensus, data} and \emph{application}~\cite{NeudeckerH19comsur}. The two-tier representation we adopt allows us to isolate the aspects that are relevant to our present focus on smart contracts and smart contract languages.} both of which influence the way smart contracts are programmed, their efficiency and the security properties they convey. Below, we outline the key aspects of the design of smart contract languages in relation to the distinguishing features of this layered architecture.  

\begin{figure}[t]
  \includesvg[width=\columnwidth]{tx-lifecycle.svg}
  \caption{\chadded{Life cycle of transactions. The blue, green and red boxes represent, respectively, the users submitting transactions, the networking nodes and the consensus nodes of a blockchain. In section $1$ of the figure, the users create transactions and transmit them to some networking node (NN); the networking nodes, in turn, run a gossiping protocol to share the knowledge of the received transactions. The mempool (dashed green container) is a distributed data structure that represents this shared knowledge of transactions. In section $2$, we see the consensus nodes (CN) collect the transactions from the mempool, propose blocks of transactions (the yellow boxes), and gossip them to the other consensus nodes. Section $3$ shows the blockchain extended with the new block selected by the consensus nodes.}}
  \label{fig:tx-lifecycle}
\end{figure}

\subsection{The Consensus layer}
The consensus layer rests on the data and network services provided by the underlying infrastructure and sets the rules for participation in the blockchain platform.
The rules vary from platform to platform, but generally include a protocol for propagating transactions across the networking nodes, and a consensus protocol for ordering the transactions and grouping them into blocks.
\chadded{%
In the transaction gossiping protocol (part $1$ of \Cref{fig:tx-lifecycle}), the networking nodes broadcast the transactions they received from users, collecting them into a distributed data structure (called \emph{mempool}).}%
\footnote{A few blockchain platforms (\eg, Hedera and IOTA) deviate from this design pattern, avoiding the transaction mempool.}
\chadded{%
In the consensus protocol (part $2$), the nodes select a set of transactions from the mempool, and order it into a \emph{block} that they propose to the other consensus nodes.
The consensus nodes then run a protocol to choose, among the proposed blocks, which one will be the next block in the sequence of blocks constructed so far --- the so-called \emph{blockchain}.
Once the consensus nodes reach an agreement on one of the proposed blocks, the chosen block is cryptographically linked to the previous ones (\eg, the new block contains the hash of the previous one), effectively making it part of the blockchain (part $3$).
}

At the consensus layer, the blockchain can be seen as a global state machine whose state (replicated at all consensus nodes) is the block\-chain, and the state transitions coincide with (the steps that contribute to) the additions of new blocks. 

\subsubsection{Key properties and incentives}
The consensus layer must guarantee three key properties: \emph{safety} (honest nodes have the same view of the blockchain), \emph{liveness} (new transactions are regularly added to the blockchain), and \mbox{\emph{finality}} (the transactions added to the blockchain are never reverted).
%
In \chdeleted{decentralized} permissionless blockchains, our focus in the present paper, these properties must be enforced without assuming any specific notion of trust among nodes, except that the majority of resources (computational or financial) is controlled by \emph{rational} nodes that participate in the protocols for profit. 
Consequently, the consensus protocols must be resistant to \emph{Sybil attacks}, making sure that artificially crafting new nodes does not give more than a negligible advantage to the adversary. Such attacks are mitigated by providing economic incentives to honest nodes that play by the rules. 
In addition to block rewards, these incentives come in the form of \keyterm{fees} that depend on various factors, \eg the amount of work needed to execute a transaction, the size of the allocated storage, and the pace at which a transaction is included in a block.
To avoid incurring higher costs than needed, developers must adopt design patterns that reduce the amount of on-line computations and on-chain storage in favour of their off-chain counterparts. 

\subsubsection{Transaction ordering}
\label{sec:overview:consensus:transaction-ordering}

Most consensus protocols leave the participant nodes free to choose which transactions from the mempool to include in a block, and in which order. 
As a result, such protocols provide no guarantee of a 
\emph{fair ordering}~\cite{Kelkar22asiapkc} on how transactions are processed. This, in turn, may open the door to attacks against contracts whose logic depends on the order in which their triggering transactions are processed: 
\eg, a user may send a transaction to reveal the solution to a bounty contract, while another user front-runs that transaction to win the bounty. 
Some blockchain platforms are systematically targeted by these attacks, which have detrimental effects on decentralization, transparency, and trustworthiness~\cite{Daian20flash,Qin21quantifying}.  
%
%
From the point of view of developers, transaction-order dependence could be mitigated, in principle, by crafting contracts so that any transaction can be executed in exactly one state. In practice, doing so would create an unacceptable congestion effect in high-bandwidth contracts, like \eg those used in DeFi.
More effective forms of mitigation are possible through ad-hoc protocols~\cite{Heimbach22aft}.

\subsection{The Contract layer}
\label{sec:overview:contract}

The contract layer sits on top of the consensus layer and  hosts the execution environment for smart contracts.
Whereas at the consensus layer we see the blockchain as a 
state machine whose state transitions correspond to the additions of new blocks,
at the contract layer what we observe is the execution of each transaction, \ie the smart contract rules it activates and their interaction with the environment. 
As a result, though smart contracts may be programmed to perform arbitrary tasks, especially in Turing-complete languages, at the contract layer the blockchain is best understood as an asset-exchange state machine in which the state keeps track of the asset balance for each user, and every transaction contributes to a state transition by either creating new assets or exchanging existing assets among users. Smart contracts and smart contract languages may be classified accordingly, based on the model they adopt for representing the balance state and the accounting of assets.

\subsubsection{Accounting models}
\label{sec:overview:compute:account-vs-utxo}

Two main models have emerged so far: \emph{account-based} and \emph{UTXO} models.%
\footnote{Given that there appears to be, as yet, no standard terminology for these concepts, we adopt naming schemes that we believe are best suited to render the underlying concepts and help grasp the key features of the existing platforms.
}
The former was first introduced by Ethereum and then adopted or revisited by other mainstream blockchains, including \eg Solana, Avalanche C-Chain, Aptos, Hedera, Algorand and Tezos. The latter was introduced by Bitcoin, and then extended by Cardano and IOTA. 

\paragraph{Account-based model} 
In the account-based model, the block\-chain state stores the deployed contracts and keeps track of the asset balance (henceforth the \emph{balance state}) as a map that associates each account with the amount of assets the account owns. 
Accounts come in two types: user accounts
and contract accounts,  each equipped with a balance. 
Transactions update the balance state by either deploying a new (user or contract) account 
or changing the account-balance map: an asset-transfer transaction is enabled only if the sender account owns all the assets to be transferred. 
In general, in the account-based model a transaction specifies
\begin{inlinelist}
\item the users who have signed the transaction,
\item the receiver account (in case it is a contract account, the transaction includes the function to be invoked and its arguments),
\item the amounts of assets to be transferred from the signers to the receiver,
and 
\item the transaction fee.
\end{inlinelist}

To illustrate, a \contract{Bank} contract handling deposits and withdraws would be structured as in the following pseudo-code: 

\begin{lstlisting}[language=pseudocode,morekeywords={Bank,deposit,withdraw,getTotalBalance}]
    contract Bank {
      var accounts         // map (user => asset) 
      deposit() { 
        expect [k]=tx.signed // tx is signed by k
        v = tx.from(k) // tokens sent to contract
        accounts[k]+=v          // trace transfer
      }
      withdraw(amnt) {
        expect [k]=tx.signed
        require accounts[k]>=amnt
        send(amnt,k)            // transfer to k
        accounts[k]-=amnt       // trace transfer
      }
      getTotalBalance() {
        return balance        // contract balance 
      }
    }
\end{lstlisting}
The contract uses the local (persistent) \code{accounts} variable, a key-value map that keeps track of the amounts deposited and withdrawn: this map provides the code-level representation corresponding to the underlying cryptocurrency balances associated with the \contract{Bank} contract and its users' accounts.   
Once the contract is created, the \txcode{deposit} and \txcode{withdraw} actions operate at two levels: on the \code{accounts} map and on the underlying balances of the accounts involved in the transaction. The \code{accounts} map is updated explicitly by the contract code, while the underlying balances are updated implicitly by the runtime. 
A user $\pmvA$ willing to deposit cryptocurrency tokens may do so by signing a transaction with receiver $\txcode{Bank}$ that invokes $\txcode{deposit}$. Executing the action checks that the amount is authorized by the signer, \emph{automatically} subtracts the specified amount of tokens from $\pmvA$'s account (assuming that there are enough), and adds an equal amount to the balance of the $\txcode{Bank}$ account (noted \code{balance} in the pseudo-code).
The $\txcode{withdraw}$ method, in turn, allows anyone to withdraw from the $\txcode{Bank}$, provided that they have previously deposited enough tokens. If so, \mbox{\solkeyword{send}\code{(amnt,k)}} removes \code{amnt} tokens from the contract balance and adds them to the balance of the signer's account, updating the \code{accounts} map accordingly. 

\paragraph{UTXO model} 
Unlike the account-based model, the UTXO model makes no reference to any explicit notion of account balance in its representation of the contract-level blockchain state. Instead, the balance for each user is traced implicitly by the  \textit{inputs} and \textit{outputs} carried along within the executed transactions. Transaction \emph{outputs} include an amount of assets and a script that specifies the \textit{spending condition} for these assets, \ie the condition stating how they can be \emph{redeemed by} (\ie unlocked to be transferred to) another transaction. 
Transaction \emph{inputs}, in turn, are references to unspent (\ie yet to be transferred) outputs of previous transactions, and provide data and the \emph{witnesses} to \textit{unlock} (\ie validate) the spending conditions of the referenced output. 
In other words, each new transaction spends outputs of previous transactions, and produces new outputs that can be consumed by future transactions. 
Each unspent output can only be consumed once, as a whole, by exactly one input. 
Then, the blockchain state is encoded as the set of unspent transaction outputs: the balance for each user is the sum of all the unspent outputs within the transactions that can be redeemed by the user (\ie those directed to the public keys the user controls). 

\newcommand{\txMiniA}{
  \scalebox{0.6}{
    \hspace{-5pt}
      \begin{tabular}[t]{|l|}
      \hline
      \\[-9pt]
      \multicolumn{1}{|c|}{$\txT[1]$} \\[1pt]
      \hline
      $\cdots$ \\[1pt]
      \hline
      \txOut[0]{\,} \\[1pt]
      \hspace{5pt} $\txscript = \pmvA\;\code{in}\;\rtx.\txSigned{}$ \\[1pt]
      \hspace{5pt} $\txval = 1:\tokT$ \\[1pt]
      \hline
    \end{tabular}}
}

\newcommand{\txMiniB}{
  \scalebox{0.6}{
    \hspace{-5pt}    
      \begin{tabular}[t]{|l|}
      \hline
      \\[-9pt]
      \multicolumn{1}{|c|}{$\txT[2]$} \\[1pt]
      \hline
      \txIn[0]{\,} \\[1pt]
      \hspace{5pt} $\txOut{} = \txT[1].\txOut[0]{}$    
      \\[1pt]
      \hline
      $\txSigned{} = [\pmvA]$ \\[1pt]
      \hline            
      \txOut[0]{\,} \\[1pt]
      \hspace{5pt} $\txscript = \code{owner}\;\code{in}\;\rtx.\txSigned{} \; \code{and}$
                      $\rtx.\txOut[0]{}.\txscript == \txscript \; \code{and}$ 
                      $\rtx.\txOut[0]{}.\txval == \txval$
                     \\[1pt]
      \hspace{5pt} $\txstorage = \{ \code{owner}:\pmvA \}$ \\[1pt]
      \hspace{5pt} $\txval = 1:\tokT$ \\[1pt]
      \hline
    \end{tabular}}
}

\tikzstyle{tx} = [rectangle, minimum width=0cm, minimum height=1cm,text centered, draw=black, fill=gray!10, draw=white, inner sep=0, outer sep=0]

To illustrate, the transaction below has a single unspent output holding one token, noted $1:\tokT$. Its script requires the redeeming transaction ($\rtx$) to include $\pmvA$'s signature in its signers list:
\begin{center}
  \begin{tikzpicture}[node distance=2cm]
    \node (tx1) [tx] {\txMiniA};
  \end{tikzpicture}
\end{center}

To spend $\txT[1]$'s single output (noted $\txT[1].\txOut[0]{}$), a redeeming transaction must refer to $\txT[1]$ from its inputs and validate the script by having $\pmvA$ as (one of) its signers. This is accomplished by the transaction $\txT[2]$ below: 
\begin{center}
  \begin{tikzpicture}[node distance=2cm]
    \node (tx2) [tx] {\txMiniB};
  \end{tikzpicture}
\end{center}

$\txT[2]$ is signed by $\pmvA$ and its script checks that
\begin{inlinelist}
\item the redeeming transaction is signed by the user stored in the \code{owner} field of the current transaction;
\item the script and the value in the redeeming transaction are the same as in $\txT[2]$.
\end{inlinelist}
Note that, although any transaction redeeming $\txT[2]$ must preserve its script and value, it can change the \code{owner}. 
In a sense, the script implements a non-fungible token (NFT): to change the ownership of the NFT, the current owner must spend the output with a new transaction (signed by herself) that specifies the new owner.

\paragraph{Account-based \emph{vs.} UTXO}
We can compare the two accounting models along two main dimensions: (i) the design patterns induced by their representation of the contract-level state, and (ii) their interaction with the underlying consensus protocols. 

At the design level, the account-based model is typically perceived as more intuitive and friendly, as it rests on programming concepts that are familiar to developers. 
Simply, contracts are standalone modules collecting executable services to be invoked by the users via transactions that operate on the assets kept in their accounts.
In the UTXO model, instead, assets and contracts are interdependent, the latter acting as guards for the former, both embedded within transactions with no explicit reference to any notion of user account.   
In the simplest incarnations of the UTXO model (such as the one in the previous example), each unspent output is managed by the associated script in the transaction.
More complex scripts are also at the avail of programmer, to
express transactions that consume multiple unspent outputs and create multiple new outputs. Still, the resulting programming practice remains somewhat cumbersome (cf.\ \Cref{sec:overview:bet} for a comparison on a concrete example, and \Cref{sec:comparison:programming-style} for a discussion of actual smart contract languages). 

As to the interaction with the underlying consensus layer, the two models have trade-offs. On the one hand, UTXO models are  exposed to liveness failures, as triggering a transaction may get stuck because all the referenced  UTXOs are spent by other transactions. The resulting \href{https://plutus-apps.readthedocs.io/en/stable/plutus/howtos/writing-a-scalable-app.html#utxo-congestion}{UTXO congestion} effect, occurring when multiple transactions try to spend the same output, \chadded{represents a non-trivial challenge for developers, especially for high-bandwidth contracts such as, \eg, Decentralized Finance (DeFi) protocols~\cite{cardano-scalable-plutus,sundae-concurrency}.}
On the other hand, the account-based model appears weaker in that it is exposed to transaction-ordering attacks. As we said earlier (cf. \Cref{sec:overview:consensus:transaction-ordering}), given that the balance state is updated only when transactions are committed,  
account-based models leave transaction senders with no means to predict whether, when and in which balance state their transactions are executed. The resulting effect, known as \emph{transaction-ordering dependence}, is troublesome as it opens the door to a variety of security attacks~\cite{ABC17post}. In blockchains where the consensus protocol does not guarantee fair transaction ordering, 
such attacks are carried out systematically by colluding consensus nodes, which leverage the economic incentives of contracts to extract value from user transactions \cite{Daian20flash,Qin21quantifying}. A further class of attacks exploit the dependency on transaction ordering to alter the contract execution flow and, consequently, the transaction fees.  
In the UTXO model, instead, a transaction can be executed in exactly one state, given by the UTXOs in its inputs. As a result, UTXO scripts do not have any dependency on transaction ordering, nor do they incur in attacks exploiting transaction fees. 

\subsubsection{Contract storage models} 
While the choice of the accounting model is certainly the classification dimension for contract languages, another aspect that is worth emphasizing is the way that smart contract languages account for a notion of \textit{persistent} contract state.  
Smart contracts often require some kind of memory to keep track of data and information that should persist across multiple executions. In programming language jargon that memory would be called \textit{state}, but to avoid confusion with other notions of block\-chain state we refer to it as (\textit{contract}) \textit{storage}. Account-based models typically encompass \emph{stateful} contracts, which encapsulate the storage directly with themselves as the \code{accounts} map in the \contract{Bank} contract. Notable exceptions are \href{https://aptos.dev/aptos-white-paper}{Aptos} and \href{https://github.com/solana-labs/whitepaper/blob/master/solana-whitepaper-en.pdf}{Solana}, in which the contract storage is held in separate data structures (\eg, accounts) and referenced from the contract. In UTXO models, instead, contracts are by default \emph{stateless} scripts that are discarded 
once their associated output is spent. A stateful form of UTXO contract may still be accounted for, however, by using the transaction fields as storage, and requiring the spending and redeeming transactions to contain the same contract with updated data in transaction fields (cf. the UTXO version of the bet contract in \Cref{sec:overview:bet})

\subsection{Exemplifying smart contracts at work: a \textit{bet} contract}
\label{sec:overview:bet}
We conclude this overview with a more extensive example that illustrates the different design patterns in the account-based and UTXO models. We use again pseudo-code to show the core concepts and stay away from the specific features of the different blockchains and contract languages. 

The contract involves two players who can join the bet by depositing 1 unit of token~$\tokT$ each. When the players join, they choose an oracle who will determine the winner, and set a deadline to close the bet 1000 blocks after the one where the join occurred, at the latest. When the oracle announces the winner, the winner can redeem the whole pot of $2:\tokT$; if instead the oracle does not choose the winner by the deadline, then both players can redeem 
their bets, withdrawing $1:\tokT$ each. 

\Cref{fig:bet:acct-stateful} shows the stateful version of the account-based contract. The stateless version in \Cref{fig:bet:acct-stateless}  follows the same design with the difference that the contract variables must be stored in a separate account, owned by the contract, and accessed from within the contract with a reference to that account that is passed as an argument to all the contract methods. 

\begin{figure}[t!]
  \begin{lstlisting}[language=pseudocode,morekeywords={STFUL_ACCT_Bet,join,win,timeout}]
    contract STFUL_ACCT_Bet  { 
      var p1,p2,oracle,deadline; // storage  
      join(o) {
        require balance==0:T  // init condition
        expect [k1,k2]=tx.signed
        require tx.from(k1)==1:T // get 1:T from k1 
             && tx.from(k2)==1:T // get 1:T from k2 
        // at this point, balance==2:T
        p1=k1; p2=k2; oracle=o
        deadline=blockH+1000 // block height + 1000
      }
      win(winner) {
        expect [o]=tx.signed
        require o==oracle && balance==2:T
        require winner==p1 || winner==p2
        send(2:T,winner)
      }
      timeout() {
        require blockH>deadline && balance==2:T
        send(1:T,p1); send(1:T,p2)
      }    
  \end{lstlisting}
  \negcaptionspace
  \caption{Account-based contract: \emph{stateful code}. 
  The players start the contract by calling \smalltxcode{join}, which requires them to deposit $1:\tokT$ each and to set an oracle. 
  The first condition ensures that \smalltxcode{join} is the first action triggered: the (system controlled) variable \smallcode{balance} is initialized to $0$ and automatically updated by the transaction (referenced to by $\smallcode{tx}$) invoking the \smalltxcode{join} method. 
  The \smallcode{expect} clause requires that the transaction is signed by exactly two keys, and binds them to \smallcode{k1} and \smallcode{k2}.  
  The next condition requires that each player deposits $1:\tokT$ 
  in the contract along with the call: namely, executing \smalltxcode{join}
  removes $1:\tokT$ from the accounts of both players,
  and adds $2:\tokT$ to the contract balance.
  Finally, the players and the oracle identifiers are recorded in the contract storage together with the deadline.
  The \smalltxcode{win} action transfers $2:\tokT$ to the winner,
  chosen between the two players by the oracle,
  who is the only possible caller. 
  Both players can call \smalltxcode{timeout} after the deadline to redeem their bets.
  }
  \label{fig:bet:acct-stateful}
  \vspace{0pt}
\end{figure}

\begin{figure}[t!]
  \begin{lstlisting}[language=pseudocode,morekeywords={STLESS_ACCT_Bet,join,win,timeout}]
    contract STLESS_ACCT_Bet { 
      join(s,o) {
        require owns(STLESS_ACCT_Bet,s)
        require s.balance==0:T
        expect [k1,k2]=tx.signed
        send(k1,1:T,s); send(k2,1:T,s)
        s.p1=k1; s.p2=k2; s.oracle=o
        s.deadline=blockH+1000
      }
     win(s,winner) {
        require owns(STLESS_ACCT_Bet,s)
        expect [o]=tx.signed
        require o==s.oracle && s.balance==2:T
        require winner==s.p1 || winner==s.p2   
        send(s,2:T,winner)
      }
      timeout(s) {
        require owns(STLESS_ACCT_Bet,s)
        require blockH>s.deadline && s.balance==2:T 
        send(s,1:T,s.p1); send(s,1:T,s.p2)
      }
    }
  \end{lstlisting}
  \negcaptionspace
  \caption{Account-based contract: \emph{stateless code}. 
  Each method takes an extra parameter $\code{s}$, that is an account to store the contract state: deploying an instance of the contract requires to  generate a new account to store its state. The condition \smallcode{owns(STLESS\_ACCT\_Bet,s)} ensures that the store is controlled by the contract: if not, it would be easy for an adversary to execute a contract action in an illegal state, subverting the contract rules. Unlike in the stateful version of the contract, inbound tokens are not passed along with the contract call, but are rendered as explicit \smallcode{send} actions. The owners of the accounts where these tokens are taken from must authorize the transfer, by signing the transaction (as done in the \smalltxcode{join} method).
  }  
  \vspace{-1pt} 
  \label{fig:bet:acct-stateless}
  \end{figure}


\smallskip
The UTXO contract, in~\Cref{fig:bet:utxo-tx,fig:bet:utxo-stateful}, draws on very different design principles. The $\txT[\code{init}]$ transaction constructs the contract script \contract{UTXO\_Bet}, which is then passed unchanged to $\txT[\code{tjoin}]$ along with an updated $\txcode{data}$ field.  
The transactions $\txT[\code{join}]$, $\txT[\code{win}]$ and $\txT[\code{timeout}]$ in~\Cref{fig:bet:utxo-tx}, in turn. act as the activating actions for the contract rules corresponding to the account-based methods. Finally, $\txT[\pmvA]$ and $\txT[\pmvB]$ represent the players' bets. The script \contract{UTXO\_Bet} ensures that 
\begin{inlinelist}
\item the contract is preserved when spending $\txT[\code{init}]$ with $\txT[\code{join}]$,
\item the storage is updated correctly,
\item and the terminal transactions $\txT[\code{win}]$ and $\txT[\code{timeout}]$ correctly transfer the funds from the contract to the players.
\end{inlinelist}

\newcommand{\txBetPlayer}[1]{
  \scalebox{0.6}{
    \hspace{-5pt}
      \begin{tabular}[t]{|l|}
      \hline
      \\[-9pt]
      \multicolumn{1}{|c|}{$\txT[#1]$} \\[1pt]
      \hline
      $\cdots$ \\[1pt]
      \hline
      \txOut[0]{\,} \\[1pt]
      \hspace{5pt} $\txscript = {#1}\;\code{in}\;\rtx.\txSigned{}$ \\[1pt]
      \hspace{5pt} $\txval = 1:\tokT$ \\[1pt]
      \hline
    \end{tabular}}
}

\newcommand{\txBetInit}{
  \scalebox{0.6}{
    \hspace{-5pt}    
      \begin{tabular}[t]{|l|}
      \hline
      \\[-9pt]
      \multicolumn{1}{|c|}{$\txT[\code{init}]$} \\[1pt]
      \hline
      $\cdots$ \hspace{70pt} \\[1pt]      
      \hline
      \txOut[0]{\,} \\[1pt]
      \hspace{5pt} $\txscript = \contract{UTXO\_Bet}$ \\[1pt]
      \hspace{5pt} $\txval = 0:\tokT$ \\[1pt]
      \hline
    \end{tabular}}
}

\newcommand{\txBetJoin}{
  \scalebox{0.6}{
    \hspace{-5pt}
      \begin{tabular}[t]{|l|}
      \hline
      \\[-9pt]
      \multicolumn{1}{|c|}{$\txT[\code{join}]$} \\[1pt]
      \hline
      \begin{tabular}{l|l|l}
        \hspace{-6pt}\txIn[0]{\,} & \txIn[1]{\,} & \txIn[2]{\,}
        \\[1pt]
        \hspace{-1pt} $\txOut{} = \txT[\code{init}].\txOut[0]{}$
        &
        \hspace{5pt} $\txOut{} = \txT[\pmvA].\txOut[0]{}$
        &
        \hspace{5pt} $\txOut{} = \txT[\pmvB].\txOut[0]{}$
        \\[1pt]
        \hspace{-1pt} $\txWit{} = \{ \code{op}:\txcode{"join"}, \code{p1}:\pmvA, \code{p2}:\pmvB, \code{oracle}:\pmv{O}, \code{winner}:- \}$
        &
        \hspace{5pt} $\txWit{} = \{ \}$
        &
        \hspace{5pt} $\txWit{} = \{ \}$
      \end{tabular}
      \\
      \hline      
      $\txSigned{} = [\pmvA,\pmvB]$ \\[1pt]
      \hline
      \txOut[0]{\,} \\[1pt]
      \hspace{5pt} $\txscript = \contract{UTXO\_Bet}$ \\[1pt]
      \hspace{5pt} $\txstorage = \{ \code{p1}:\pmvA, \code{p2}:\pmvB, \code{oracle}:\pmv{O}, \code{deadline}: h \}$ \\[1pt]
      \hspace{5pt} $\txval = 2:\tokT$ \\[1pt]
      \hline
    \end{tabular}}
}

\newcommand{\txBetWin}{
  \scalebox{0.6}{
    \hspace{-5pt}    
      \begin{tabular}[t]{|l|}
      \hline
      \\[-9pt]
      \multicolumn{1}{|c|}{$\txT[\code{win}]$} \\[1pt]
      \hline
      \txIn[0]{\,} \\[1pt]
      \hspace{5pt} $\txOut{} = \txT[\code{join}].\txOut[0]{}$
      \\[1pt]
      \hspace{5pt} $\txWit{} = \{ \code{op}:\txcode{"win"}, \cdots, \code{winner}:\pmvA \}$
      \\[1pt]
      \hline
      $\txSigned{} = [\pmv{O}]$ \\[1pt]
      \hline            
      \txOut[0]{\,} \\[1pt]
      \hspace{5pt} $\txscript = \pmvA\;\code{in}\;\rtx.\txSigned{}$ \\[1pt]
      \hspace{5pt} $\txval = 2:\tokT$ \\[1pt]
      \hline
    \end{tabular}}
}

\newcommand{\txBetTimeout}{
  \scalebox{0.6}{
    \hspace{-5pt}    
      \begin{tabular}{|l|}
      \hline
      \\[-9pt]
      \multicolumn{1}{|c|}{$\txT[\code{timeout}]$} \\[1pt]
      \hline
      \txIn[0]{\,} \\[1pt]
      \hspace{5pt} $\txOut{} = \txT[\code{join}].\txOut[0]{}$ \\[1pt]
      \hspace{5pt} $\txWit{} = \{ \code{op}:\txcode{"timeout"}, \cdots \}$
      \\[1pt]      
      \hline
      $\txSigned{} = []$ \\[1pt]
      \hline      
      \begin{tabular}{l|l}
        \hspace{-6pt}\txOut[0]{\,}
        &
        \txOut[1]{\,}
        \\[1pt]
        \hspace{-1pt} $\txscript = \pmvA\;\code{in}\;\rtx.\txSigned{}$ 
        &
        \hspace{5pt} $\txscript = \pmvB\;\code{in}\;\rtx.\txSigned{}$
        \hspace{-5pt}
        \\[1pt]
        \hspace{-1pt} $\txval = 1:\tokT$
        &
        \hspace{5pt} $\txval = 1:\tokT$
      \end{tabular}
      \\
      \hline
  \end{tabular}}
}

\tikzstyle{tx} = [rectangle, minimum width=0cm, minimum height=1cm,text centered, draw=black, fill=gray!10, draw=white, inner sep=0, outer sep=0]

\begin{figure}[t]
  \begin{tikzpicture}[node distance=2cm]
    \node (txA) [tx] {\txBetPlayer{\pmvA}};
    \node (init) [tx,left of=txA, xshift=-0.8cm] {\txBetInit};
    \node (txB) [tx, right of=txA, xshift=1cm] {\txBetPlayer{\pmvB}};
    \node (join) [tx, below of=txA, yshift=-0.3cm] {\txBetJoin};
    \node (win) [tx, below of=join, xshift=-2.65cm, yshift=-0.7cm] {\txBetWin};
    \node (timeout) [tx, right of=win, xshift=2.4cm, yshift=0cm] {\txBetTimeout};  
  \end{tikzpicture}
  \caption{Transactions for the UTXO-based bet contract. $\txT[\code{init}]$ creates the contract: the script \smalltxcode{UTXO\_Bet} is specified 
  in~\Cref{fig:bet:utxo-stateful}.
  Players join the contract by sending $\txT[\code{join}]$, which spends $\txT[\code{init}]$ (preserving its script), and the two $1:\tokT$ bets provided by the players.
    Note that both players must sign $\txT[\code{join}]$, and so they must agree on the values of the witnesses $\code{p1}$, $\code{p2}$ and $\code{oracle}$. 
    $\txT[\code{join}]$ records these witnesses in its storage,
    as well as the $\code{deadline}$.
    $\txT[\code{join}]$ can be spent either by $\txT[\code{win}]$ or
    $\txT[\code{timeout}]$, which terminate the contract.
    $\txT[\code{win}]$ must be signed by the oracle,
    and can be spent only by the $\code{winner}$ set in the witnesses.
    In the figure, we assume that the winner is $\pmvA$, and accordingly
    the script of $\txT[\code{win}]$ requires that the redeeming 
    transaction ($\rtx$) is signed by $\pmvA$. 
    $\txT[\code{timeout}]$ requires no signatures, and it
    splits $2:\tokT$ in two outputs of $1:\tokT$ each, 
    that can be spent by the two players.
    The script $\contract{UTXOBet}$ in $\txT[\code{join}]$ ensures that the transactions
    $\txT[\code{win}]$ and $\txT[\code{timeout}]$ are constructed according
    to these rules.
  }
  \label{fig:bet:utxo-tx}
\end{figure}

\begin{figure}[t]
  \begin{lstlisting}[language=pseudocode,morekeywords={UTXO_Bet,join,win,timeout}]
    contract UTXO_Bet { // Stateful UTXO-based
      expect [op,p1,p2,oracle,winner]=rtx.in[0].wit
      if op=="join":
        // rtx must be signed by p1,p2 to redeem 1:T
        require rtx.signed==[p1,p2]
        require value==0:T
        expect [ri0,ri1,ri2]=rtx.in    
        require ri0==ctxo // Tjoin.in[0]=Tinit.out[0]
        require ri1.value==1:T // p1's bet
        require ri2.value==1:T // p2's bet
        expect [ro0]=rtx.out   // Tjoin has 1 output        
        require ro0.script==script // preserve script
        require ro0.value==2:T
        require ro0.data.p1==p1 // set p1
        require ro0.data.p2==p2 // set p2
        require ro0.data.oracle==oracle // set oracle
        require ro0.data.deadline==blockH+1000
      elif op=="win":
        require rtx.signed==[data.oracle]
        require len rtx.in==1 // Twin has 1 input
        require winner in [data.p1,data.p2]
        expect [ro0]=rtx.out
        require ro0.script=="{winner} in rtx.signed"
        require ro0.value==2:T
      elif op=="timeout":
        require blockH>data.deadline
        require len rtx.in==1 // Ttimeout has 1 input
        expect [ro0,ro1]=rtx.out // ... and 2 outputs
        require ro0.script=="{data.p1} in rtx.signed"
        require ro0.value==1:T
        require ro1.script=="{data.p2} in rtx.signed"
        require ro1.value==1:T
      else require false
    }
  \end{lstlisting}
  \negcaptionspace
  \caption{Pseudocode of a \emph{stateful UTXO-based} bet contract. The 
  spending condition is a switch between three cases, corresponding to the transactions
  $\txT[\code{join}]$, $\txT[\code{win}]$ and $\txT[\code{timeout}]$.
  Note that only the first case requires the script to be preserved,
  while the others define the scripts of the redeeming transactions 
  as simple a signature verification, terminating the contract and transferring the funds to the players ($2:\tokT$ to the winner for $\txT[\code{win}]$, and $1:\tokT$ each for $\txT[\code{timeout}]$).}
  \label{fig:bet:utxo-stateful}
\end{figure}


\subsection{\chadded{Cross-chain interactions}}

\chadded{DApps can span across multiple blockchains, making it possible the exchange of different native crypto-assets \cite{ZamyatinAZKMKK21fc,BelchiorVGC22csur,RenHLNOTZ23tkde}. 
In general, cross-chain interactions presuppose a communication layer (\eg, a decentralized bridge systems), and a consensus-agnostic communication protocol (\eg, \href{https://chain.link/cross-chain}{CCIP} from Chainlink). Cross-chain interactions are important, but clearly out of scope for our comparison of smart contract languages. That said, a special mention is in order for {\em native} cross-chain architectures, as they may be seen as an alternative to smart contracts in the DApps paradigm. In fact, such architectures are designed to host multiple, application-specific blockchains, each tailored for a given use case, and communicating through specific protocols. In other words, having multiple blockchains each running a single contract is as an alternative to deploying multiple contracts on a single blockchain.}

\chadded{
Notable cases of native cross-chain architectures include \href{https://cosmos.network/}{Cosmos} and 
\href{https://polkadot.network/}{Polkadot}. In Cosmos, application-specific blockchains are called \textit{Appchains}, and the interaction among different contracts is rendered as inter-chain communication over the IBC protocol. The protocol manages specific operations such as the transfer of tokens both between accounts in the same Appchain and across accounts operating on different Appchains. Appchains may be programmed in GoLang (a general-purpose programming language), and 
\href{https://github.com/CosmWasm/cosmwasm/}{CosmWasm} (a Rust derivative). Communication with blockchains external to Cosmos relies on bridges that support IBC (such as \href{https://www.gravitybridge.net}{Gravity}). In Polkadot, the application-specific blockchains are called \textit{parachains}, implemented through the \href{https://substrate.io}{Substrate} framework with its native smart contract language \href{https://use.ink/}{ink!} (again a Rust derivative). Parachains communicate via the \textit{Cross-Consensus Messaging} (XCM) language over the  transport layer provided by the Polkadot network. XCM is designed to be used outside of Polkadot as well, but requires the implementation of a dedicated bridge.}

%% file: languages.tex
\section{A tour of smart contract languages}
\label{sec:languages}

We overview in this section the main features of our selection of smart contracts languages.

%% file: solidity-ethereum.tex
\subsection{Solidity / Ethereum}
\label{sec:solidity}

Solidity is one of the first contract languages, dating back to 2014, and it is  currently the main high-level contract language for the
blockchains that support the Ethereum Virtual Machine (EVM), \ie Ethereum, Avalanche C-Chain, and Hedera
among the others.
Solidity contracts must be compiled to EVM bytecode in order to be executed by the consensus nodes of these blockchains.


Solidity adheres to the account-based stateful model outlined in~\Cref{sec:overview:compute:account-vs-utxo}.
Accounts are partitioned into user accounts (a.k.a. Externally Owned Accounts, or EOAs) and contract accounts, and are uniquely identified by an \emph{address}.
Contracts, akin to classes in object-oriented languages, have methods to access and update the storage, which consists of
the contract balance and variables. 
Contract variables can record fixed-size data as well as dynamic data structures like arrays and key-value maps.
Transactions are signed by a single EOA and trigger contract calls, possibly transferring units of the native cryptocurrency (ETH) from the caller EOA to the contract. The called contract, in turns, can trigger calls to other contracts. 
Transactions can deploy new contracts;
the same contract code can be deployed multiple times, each instance having its own address and storage.
Control structures include unbounded loops and recursion, but in practice all computations are bounded by the fee mechanism
(see~\Cref{sec:comparison:fees}).

Solidity is statically typed, with 
types of variables and methods specified explicitly by the programmer.
It features subtype polymorphism and ad-hoc polymorphism.
Some types support type-safe implicit conversions;
type-unsafe explicit conversions lead to compile errors.
The language has some type-unsafe primitives 
(\eg, low-level calls and inline assembly),
which require the programmer to explicitly take care of the format of data.
Solidity supports multiple inheritance between contracts. 
Each source file can define multiple contracts and import code from external files. This allows the reuse of code components (libraries, interfaces, and contracts).

Because of its familiar JavaScript-like syntax and its procedural programming style, Solidity is usually considered an easy language to learn. However, it has a few design quirks that, together with the inherent complexity of current DApps, have deep implications on the security of smart contracts.  
We will discuss some of them later in~\Cref{sec:comparison}.

%% file: rust-solana.tex
\subsection{Rust / Solana}
\label{sec:rust-solana}

Rust is a general-purpose programming language, which was adopted as the main smart contract language for Solana.
As for Solidity on Ethereum, also Rust must be compiled to bytecode in order to be executed by Solana nodes.

Solana follows a \textit{stateless} account-based model: 
contracts take the form of procedures, without an associated state.
Therefore, any data these procedures interact with is stored within separate accounts, supplied as parameters. 
Accounts are partitioned into EOAs and contract accounts, but unlike Ethereum, in Solana
any EOA is owned by a contract account and 
can store data associated to that contract account, which instead only stores executable code and 
it is the only one with write permission.
In general, state updates are regulated by the principles of \emph{ownership} and \emph{holdership}: the entity who knows the private key is considered as the holder of the account, 
while the owner (always a contract account) is the only one that can  modify the account data.
Special pre-defined contract accounts
manage the creation of accounts, the transfer of native currency, and the minting of custom tokens. 
While this design mandates supplementary checks in the contract to ensure security, it also enables the parallel processing of transactions. 
To this purpose, transactions specify all the accounts whose data will be read or written throughout their execution: in this way, the runtime environment can detect when two transactions can be executed concurrently: namely, if no transaction reads or writes parts of the state that are written by the other transaction, then the two transactions are parallelizable~\cite{BGM21lmcs}.
Whereas in Ethereum a transaction represents a single contract call, in Solana a transaction can contain several calls, each of which may be related to a distinct contract. These calls are carried out sequentially, and the failure of any one of them results in discarding the changes of the entire transaction. The maximum size of transactions is limited to ${\sim}1\text{KB}$ 
in order to bound the amount of calls.

Rust is statically typed: notably, its type system can statically detect bugs such as null-pointer dereference, which instead lead to run-time errors in other programming languages like C++.
To do that, the type system rigorously tracks data possession throughout the program, enabling it to operate without a garbage collector by detecting memory allocations and deallocations at compile time. 
The type system ensures that references do not outlive the data they point to causing dangling pointers and that data is not mutated unexpectedly. 

Writing contracts directly in Rust poses several challenges to developers, \eg:
\begin{inlinelist}
\item contracts must be encapsulated into a single procedure, which must switch to the right part of code depending on the parameters;
\item the data structures exchanged between the contract and its clients must be manually serialized/deserialized;
\item contracts must check that the accounts passed as parameters carry the authorizations of the legitimate holders and owners (see~\Cref{sec:comparison:security}).
\end{inlinelist}
To partially relieve developers from this bureaucracy, the \href{https://www.anchor-lang.com/}{Anchor} framework offers higher-level abstractions atop the raw Rust layer~\cite{CuiZGT022ccs}. Anchor allows developers to write contracts as sets of methods, and it eliminates the need to manually encode data structures, specifying the contract interface 
through an Interface Definition Language.
Additionally, Anchor automatically performs some of the above-mentioned security checks, based on the types associated with the accounts. 
A downside is a doubling of deployment fees compared to pure-Rust.

%% file: aiken-cardano.tex
\subsection{Aiken / Cardano}
\label{sec:aiken-cardano}

Cardano is currently the main smart contract platform
following the UTXO model.
Cardano extends the UTXO model of Bitcoin in two directions~\cite{Chakravarty20wtsc}:
it follows a stateful storage model, allowing users
to include arbitrary data in transaction outputs,
and it features a Turing-complete script language,
which overcomes the expressiveness limitations of Bitcoin contracts~\cite{AtzeiBCLZ18post}.
Cardano consensus nodes execute scripts written in Plutus Core, 
a low-level untyped lambda-calculus.
Although this language is Turing-complete, in practice
computations are bounded by the fee mechanism.
There are a few high-level languages that compile into
Plutus Core, both general-purpose and DSLs.
The first high-level contract language for Cardano 
was Plutus Tx, a general-purpose 
typed functional language that is a subset of Haskell.
This allows Cardano developers to use Haskell to code both the 
on-chain and off-chain parts of a decentralized application.
A main advantage of this approach is the guarantee of 
consistency between the two parts, \eg a client will never pass
values with the wrong type to the contract.
A disadvantage is that, when one is only interested in the 
on-chain part, 
using this general framework may be unnecessarily complex.
Other languages supported by Cardano are focussed just on the on-chain part: they include \href{https://marlowe.iohk.io/}{Marlowe}
(a domain-specific language for financial contracts),
\href{https://github.com/OpShin/opshin/tree/main}{Opshin},
and \href{https://aiken-lang.org/}{Aiken}.

Aiken, in particular, is a high-level language with a minimal set of features for programming the on-chain part~\cite{Rosa23aiken},
Similarly to Plutus Tx, Aiken is a functional language
compiling to Plutus Core.
Aiken is used to write the spending conditions of UTXO transactions, akin to the pseudo-code in~\Cref{fig:bet:utxo-stateful}. 
This involves checking all the parts of the transaction output
that is being spent, and parts of the spending transaction, including its outputs, to ensure that the spending transaction represents a valid update of the contract state.
Consequently, programming in Aiken (or any other Cardano languages) requires a paradigm shift \wrt the other languages in our selection, which instead  support the procedural style. 
This has repercussions on code readability and security (see~\Cref{sec:comparison:programming-style,sec:comparison:loc,sec:comparison:security}).
%
Said that, the language is strongly typed, featuring 
algebraic types and pattern matching,
parametric polymorphism, and recursive types.
Aiken features recursion, so preserving the (theoretical) 
Turing-completeness of the underlying Plutus Core language. 
%
Aiken also supports anonymous and higher-order functions.

%% file: pyteal-algorand.tex
\subsection{(Py)TEAL / Algorand}
\label{sec:pyteal-algorand}

Algorand is a blockchain platform launched in 2019, which over the years has updated its smart contract capabilities several times, passing from a simple model of stateless contracts to Turing-powerful stateful contracts. 

Algorand follows the stateful account-based model. Every account (both user and contract) holds a balance of the native cryptocurrency and of custom tokens, as well as data associated to contracts. Unlike Ethereum, where the contract state is entirely stored in a contract account, in Algorand it is distributed across different components: a key-value storage associated to the contract account, a key-value storage associated to user accounts, and further keyed storage segments (called \emph{boxes}), used to overcome the strict size limits of the contract storage (just 8KB shared among a maximum of 64 key-value pairs).  

The Algorand nodes execute a custom bytecode, which is the compilation target of higher-level contract languages, using TEAL as an intermediate assembly-like language. 
%
The TEAL instruction set is similar to that of a stack-based machine, with only a few abstractions over low-level details. \Eg, function invocations are performed via a \emph{call} instruction rather than a plain jump, and a separate call stack is used to store function arguments and return values. Although this requires some stack manipulation to move arguments from the call stack to the operand stack whenever needed,
one can easily recover function arguments at constant offsets in the call stack, rather than having them buried deep in the operand stack. TEAL types are limited to byte arrays and unsigned integers. The contract itself is also able, when called, to generate so-called ``inner'' transactions,  which can transfer assets, call other contracts, and more.
%

To reduce the burden of directly writing TEAL bytecode, a few higher-level languages and frameworks have been proposed, \eg PyTeal, Beaker, Tealish, TealScript, and PuyaPy. Among them, the most widespread is the pairing PyTeal/Beaker, a library of Python bindings through which one can write Python code that produces TEAL bytecode at run-time. In this way, programmers can use familiar higher-level constructs, like logical/arithmetic expressions, control flow, variables and key-value maps, and functions.
Overall, the resulting code is not too dissimilar from the procedural-style code one could obtain \eg in Solidity. Still, some quirks remain about the handling of storage and of inner transactions (see~\Cref{sec:comparison:programming-style}).

%% file: move-aptos.tex
\subsection{Move / Aptos}
\label{sec:move-aptos}

Move is a smart contract language inspired by Rust that has been embedded into multiple blockchain platforms, including \href{https://diem-developers-components.netlify.app/papers/the-diem-blockchain/2020-05-26.pdf}{Libra/Diem}, 
\href{https://starcoin.org/downloads/Starcoin_Whitepaper.pdf}{Starcoin}, 
\href{https://aptos.dev/aptos-white-paper}{Aptos}
and 
\href{https://blog.sui.io/why-we-created-sui-move/}{Sui}.
One of Move's highlights is its static type system based on \emph{linear types}.
Linear types enforce the so-called \emph{must-move} semantics, ensuring that tokens (and resources in general) are never replicated or lost.
This is a major constraint when writing programs and has a number of implications on the safety properties of the compiled code.
Even though linear typing does not prevent a programmer from writing a wrong program in one way or another, it surely helps in crafting correct implementations where illicit replication or deletion of tokens is statically rejected.

Another highlight of Move is allegedly being chain-agnostic.
This is not entirely true though: each embedding must deliver a porting of the language tailored to the platform's peculiarities, providing a custom framework and a standard library, as well as applying a few tweaks to the language.
In this section we delve into Aptos, a direct successor of Libra/Diem (now dismissed).

The Move/Aptos programming model revolves around a few key principles. 
Contracts take the form of modules, containing \lstinline{struct} definitions and functions. 
Structs are the basis for representing data structures, while functions establish the only interface for module clients to create, access, or modify such data structures.
Struct fields can be accessed only from within the module code, granting information hiding and comprehensive control over the operations involving the datatypes therein defined.
Once created by a module function, the type system treats structs as first-class resources that cannot be copied or implicitly discarded, only permitting either movement between program storage locations or passing around between function calls.
This discipline takes place fully at compile time and is enforced by linear types.
Linearity checks can be disabled through \emph{abilities}: tagging a struct with the \lstinline{copy} ability renders it a \emph{value} open to duplication, while the \lstinline{drop} ability enables destruction at the end of the scope.

Different Move variants offer distinct persistent storage representations, aligning with the peculiarities of the underlying platform.
Aptos defines the \emph{global storage} as a map from account addresses to resources encoded by a struct datatype. 
The creation of a resource in the global storage is exclusive to the contract signer, performed through a special language primitive.
Accessing and modifying resources is less restrictive: anyone can request (\lstinline{borrow}) a reference to a resource via the account address under which the resource is stored. 


Move/Aptos follow a \emph{stateless} account-based model.
Global variables are not allowed unless constant, which implies that modules are stateless at the language level.
This affects how contracts are implemented: all the relevant data, say the contract \emph{state}, must be stored using user-defined datatypes and eventually retrieved from the global storage.

Each time a contract is run,
the address of the invoking user account (the \emph{signer} account) is passed as an argument with a special type \lstinline{signer} that guarantees that it
is non-copiable and it cannot be put into a user-defined struct datatype or saved on the global storage.
This design prevents a contract from performing actions on behalf of other users than the current signer. 
Such security measures have some drawbacks: for the same reason why only the signer can write a new record on the global storage, any contract involving multiple participants (\eg, auctions, bets, games, \etc) must rely on explicit \emph{opt-in}, implying a voluntary choice to engage in a specific activity.
This means that each participant has to perform the first write operation; then any participant can access data stored by other accounts through reference borrowing.

%% file: smartpy-tezos.tex
\subsection{SmartPy / Tezos}
\label{sec:smartpy-tezos}


The Tezos blockchain features a few high-level contract languages, including  
\href{https://liquidity-lang.org/}{Liquidity},
\href{https://archetype-lang.org/docs/introduction/}{Archetype},
\href{https://ligolang.org/docs/intro/introduction/?lang=jsligo}{LIGO}, 
and
\href{https://smartpy.io/}{SmartPy}.
%
Among them, the last two seem the most actively supported: here we consider SmartPy, since its Python-like style lends itself to a more direct comparison with Algorand's PyTeal. 

Tezos follows the account-based stateful model. Its consensus nodes execute low-level code written in Michelson, a statically-typed, stack-based and Turing-complete bytecode language.
SmartPy, as the other Tezos high-level languages, must be compiled into Michelson in order to be executed.


SmartPy exploits meta-programming on top of Python: \ie, SmartPy contracts are just (decorated) Python programs, which are transformed into Michelson code by the SmartPy compiler.
Meta-programming allows developers to use the syntax and control structures of SmartPy match Python's, as well as to use Python libraries. 
The language is fully typed, with type inference performed after a transformation into an intermediate OCaml code  (see~\Cref{sec:comparision:types}).
When unable to infer a datatype, the SmartPy compiler generates an error and requires an explicit cast.
Meta-programming decorators are used to specify the contract interface, \ie the set of its public functions, the contract storage, and testing scenarios.
Datatypes of the contract storage do not correspond to the native Python datatypes, but are defined through the SmartPy library.
The deployment of a SmartPy contract specifies the initial contract storage, which is set via the contract constructor. Unlike Ethereuum, this initial storage is statically incorporated in the Michelson code, and the contract cannot use external data (\eg, the caller's address) to initialize its storage. 
%
%
%
%
Contract code cannot contain externally defined data, such as  externally-defined contracts.

%% file: comparison.tex
\section{Comparative analysis}
\label{sec:comparison}

\input{tab-overview.tex}

In this section we perform a comparative analysis of the smart contract languages presented in~\Cref{sec:languages}. 
We outline below the key elements of our comparison. 
A first, high-level view is in~\Cref{tab:overview}, which classifies languages/platforms according to the architectural aspects discussed in~\Cref{sec:overview}. 
A more in-depth comparison is based on our hands-on experience on developing a common benchmark of use cases. 
We describe our benchmark in~\Cref{sec:comparison:benchmark},
and then in~\Cref{sec:comparison:programming-style,sec:comparison:loc,sec:comparison:security} we exploit it to compare the programming styles of contract languages, their verbosity and readability, and the security implications of their design.
Then, in~\Cref{sec:comparision:types,sec:comparision:verification} we discuss the role of the tool chain (compiler and static analyzers) in preventing vulnerabilities and other loopholes. 
In~\Cref{sec:comparision:onchain-offchain} we analyse the support for the integration of on-chain and off-chain components.  
In~\Cref{sec:comparison:fees} we compare the fee models of the blockchain platforms.
Finally, in~\Cref{sec:comparison:functionalities} we reflect on our experience in developing the benchmark, 
by discussing how the availability of platform functionalities affects the development of smart contracts.
\Cref{tab:strengths} summarises our assessment.



\subsection{\chadded{Smart contracts benchmark}}
\label{sec:comparison:benchmark}

\chadded{%
The ``Rosetta Smart Contracts'' benchmark~\cite{smart-contracts-comparison} is a specialization of \href{https://rosettacode.org/wiki/Rosetta_Code}{Rosetta Code} to the realm of smart contracts.
It showcases the contract languages discussed in~\Cref{sec:languages}, using them to implement a diversified class of use cases.
Two main drivers have influenced our choice of the use cases: first, to provide a representative selection of common DApp use cases, such as those in the \href{https://github.com/OpenZeppelin/openzeppelin-contracts}{Openzeppelin library} for Ethereum; secondly, to serve as an adequate test-bed for a comparative analysis of the functionalities supported by the different smart contract languages and platforms.
}

\chadded{%
The benchmark currently includes \numusecases use cases, whose implementations are distributed across 151 source code files, with a cumulative size of $\sim$900KB and $\sim$18K LoC.
%
\Cref{tab:functionalities-vs-usecases} enumerates the use cases and the functionalities required to implement them.
These functionalities represent the basic features that are provided by smart contract languages, possibly exploiting the low-level primigtives  made available by the underlying blockchain platforms where the smart contracts are executed. To illustrate, 
the Bet use case described in~\Cref{sec:comparison} requires the following functionalities:} 
\begin{inlinelist}
\item \chadded{``native tokens'': the contract involves transfers of native cryptocurrency (from the players to the contract for the \smalltxcode{join} action, and for the contract to the players for the \smalltxcode{win} and \smalltxcode{timeout} actions);}
\item \chadded{``multisig transactions'':  the \smalltxcode{join} action must be simultaneously authorized by both players;}
\item \chadded{``time constraints'': the \smalltxcode{timeout} action must be enabled after a given deadline;}
\item \chadded{``transaction revert'': some transactions must be reverted when some conditions are not satisfied (\eg, when the \smalltxcode{win} action is not authorized by the oracle).}
\end{inlinelist}  
\chadded{%
As shown later in~\Cref{tab:functionalities-vs-workarounds}, not all languages/platforms provide native support for all the functionalities listed in \Cref{tab:functionalities-vs-usecases}. When that is the case, we resort to workarounds, possibly adapting the specification of the use case (\eg, if multisig transactions are not available, in the Bet contract we can split the \smalltxcode{join} action in two actions, one for each player). See~\Cref{sec:comparison:functionalities} for more details about these workarounds.
}

\input{tab-usecases.tex}


\subsection{Comparison overview}
\label{sec:comparison:programming-style}

Roughly, we can partition the smart contract languages presented in~\Cref{sec:languages} into two classes, according to the programming style they induce: the \emph{procedural style} and the \emph{approval style}.
The former class includes languages where the contract reacts to transactions by updating its state and/or the ledger state (possibly distributed across multiple accounts): Solidity, Rust, Move and SmartPy all belong to this class.
The latter includes languages where the contract is expected to approve or discard a single transaction or a group of transactions: Aiken belongs to this class.
TEAL/PyTeal follows a hybrid approach, supporting both styles. 
As we will see in~\Cref{sec:comparison:security}, the programming style is one of the factors that contribute to the security of contracts, and it is strictly related to the level of abstraction provided by the language over the underlying blockchain platform.

\textbf{Solidity} and \textbf{SmartPy} are those that most closely follow the procedural style: contracts have code (a set of procedures) and a state that can be updated in reaction to procedure calls. Despite the strong similarity between these two languages, important differences exist. A notable one lies in the interaction with other contracts. In Solidity, a method \txcode{f} can call another contract's method \txcode{g} at any point, interrupting the execution of \txcode{f} to start that of \txcode{g}.
In SmartPy, instead, the execution of \txcode{g} takes place only \emph{after} the caller \txcode{f} has completed.
This design choice has repercussions on the programming style and on the security: on the one hand, Solidity's design leads to more natural implementations (\eg, \txcode{f} calls an oracle \txcode{g} to get some value, and then uses that value in its continuation), but on the other hand it is a cause of attacks (see~\Cref{sec:comparison:security}). 
Programming the same behaviour in SmartPy requires \txcode{g} to callback the contract of \txcode{f} after it has finished its execution, and store the return value in the caller's storage
(which is somehow less natural).

\textbf{Rust}, either raw or using \textbf{Anchor}, while still adhering to the procedural style, substantially departs from Solidity and SmartPy. This is only in part explained by the stateless nature of Solana, and by the additional checks on the accounts passed as parameters that this model requires. 
Programming in raw Rust, as discussed in~\Cref{sec:rust-solana},
requires a careful and often verbose approach, increasing error-proneness due to the extensive use of boilerplate code.
Although these issues are partially mitigated by the Anchor framework, Solana contracts are more verbose than those in other account-based platforms (see~\Cref{sec:comparison:loc}). 

\textbf{Move}, although still based on the stateless account-based model, induces a unique procedural programming style, centered around linear types.
Defining data types requires some care: assets cannot be mixed with ordinary data within the same struct, since a different treatment is needed. 
While integer values can be modified and updated like in common programming languages, assets and resources in general cannot be modified, copied or dropped and must be put into a non-copiable and non-droppable wrapper type in order to be manipulated. 
According to our experience, making a Move program compile and work properly can require a substantial effort, all the more so when dealing with asset transfers: its strict type system puts the developer on rigid rails; escaping such rails would likely lead to compile-time errors.
Similarly to Move, Rust also supports ownership types (so enforcing the same strict discipline over the copiability of datatypes).
However the programming model imposed by Solana 
does not fully exploit the Rust's complex type system, sticking to 
a more conventional programming practice.
In particular, currency and assets are not represented by uncopiable/undroppable datatypes in Rust/Solana, and so their linearity is not statically guaranteed by Rust's type system, but by run-time checks.

\textbf{Cardano} substantially differs from all the other platforms discussed in this paper, being the only representative of the UTXO model.
First, the current contract state is recorded in the current unspent transactions that encode the contract.
Then, performing a contract action means spending that transaction with a new one that sets the new contract state: therefore, the contract does not compute the new state (as in account-based platforms), but it just verifies that the state in the redeeming transaction is a correct update of the old one.
This motivates the paradigm switch from the procedural style to the approval style.
\textbf{Aiken} brings a purely functional flavour to the table, making code overall robust thanks to strong types and data immutability, albeit verbose and difficult to write for developers trained in procedural programming paradigms.
As noted in the pseudo-code of the UTXO Bet contract in~\Cref{fig:bet:utxo-stateful}, the contract script must check several transaction fields, \eg the data fields where the contract state is stored. For instance, transferring a token from the contract to some address requires checking that the spending transaction has some outputs with suitable signature verification scripts.
This workflow is more complex and verbose than in the account-based model, where an explicit call to some transfer primitive achieves the same goal. 
Admittedly, Aiken features the typical arsenal of constructs provided by functional languages, 
including the record update syntax, which somewhat reduces possible errors when updating the state.
However, when the contract logic is complex, correct state management turns out to be a cumbersome task and programmers may still introduce errors despite the robust and type-safe design of Aiken.
Unlike in account-based models, where interactions between contracts can be rendered directly as contract calls, in the UTXO model contract calls are not meaningful. Indeed, calling a contract would require the caller to perform a sort of ``internal'' transaction to trigger a computation step of the callee. 
Although these internal transactions are not featured by Cardano, 
some forms of composability between contracts are possible, \eg by multi-input transactions that force dependencies between the scripts of the spent outputs. More sophisticated interactions can be obtained by resorting to layer-2 implementations of asynchronous message-passing~\cite{Vinogradova24wtsc}.



\textbf{Algorand}, being the platform whose contract layer and languages have changed the most during its lifespan, is also the one for which it is most difficult to bring a definitive assessment. 
%
Originally, Algorand only supported \emph{smart signatures}, \ie simple stateless contracts whose primary purpose was that of deciding whether to approve the transactions coming from the smart signature's address \cite{BartolettiBLSZ21fc}. According to our rough taxonomy, smart signatures follow the approval style. 
After a number of updates, the contract layer was enriched with so-called \emph{applications}, a basic form of stateful smart contracts, but still leveraging smart signatures for handling assets transfers. This contract model was a hybrid between the approval style (needed to write the smart signature part of the contract) and procedural style (needed for the application, which handles the contract state). 
The introduction of inner transactions and application accounts (see~\Cref{sec:pyteal-algorand}) to the contract layer has made it possible to eliminate the need for smart signatures in stateful contracts, allowing them to construct and submit their own transactions. Effectively, this makes the current programming practice of Algorand adhere to the procedural style.


\subsection{Code verbosity and readability}
\label{sec:comparison:loc}

As a rough comparison between contract languages, we measure in~\Cref{tab:loc-comparison} the LoC of the implementations in our benchmark (restricting to the use cases where all the implementations are available).
As expected after~\Cref{sec:overview:bet}, the UTXO-based model, here represented by \textbf{Aiken}, leads to more verbose implementations than account-based models. 
Among the latter, \textbf{Anchor} for Rust is definitely the more verbose. This is due in part to the language bureaucracy and in part due to the need to handle data in multiple accounts, which is a consequence of how Solana renders the stateless model.
However, statelessness alone does not cause verbosity: \eg, \textbf{Move} contracts are more concise than Solana's, which is penalized by the additional account validation checks. 
The other languages in the account-based model have, on average, similar verbosity: we just note that the slightly higher LoCs of Move are counter-balanced by the increased robustness due to static typing (cf.~\Cref{sec:comparision:types}).
%
%

Regarding readability, in the absence of a widely accepted metric we resort to a qualitative evaluation. In general, we have a poor readability when understanding the behaviour of a contract requires a low-level knowledge of the structure of blockchain transactions.
This is the case \eg of Aiken and \textbf{PyTeal}: in the first case the problem seems inherent to the closeness of Aiken to the UTXO model, while in the second case it seems related to the handling of storage and of inner transactions.
PyTeal is also a witness of the fact that a good readability is not always implied by a low verbosity.  
The readability of Move contracts is strictly related to the understanding of linear types: 
developers unfamiliar with these concepts will find it quite difficult to make some sense of a Move contract. 
In Anchor/Solana, poor readability is caused by a combination of factors: unfamiliarity with the Rust ownership model and the distribution of the state across multiple accounts.

\input{tab-loc.tex}


\subsection{Security implications of language design}
\label{sec:comparison:security}

The design of a smart contract language and of the underlying contract layer has deep implications on the security of the applications built on them.
A paradigmatic example is the famous reentrancy issue of \textbf{Ethereum}, which has been the basis of several real-world attacks~\cite{AtzeiBCLZ18post,Luu16ccs}.
The issue arises from the combination of a few unfortunate design choices at the EVM level: 
\begin{inlinelist}
\item called methods always have a reference to the caller; 
\item any method can call any other method;
\item there are no bounds on the depth of nested calls;
\item the most critical contract field, the ETH balance, is implicitly updated as a side effect of method calls.  
\end{inlinelist}
Putting it all together, it may happen that when a contract calls another contract, the callee might call back its caller in such a way as to modify its state variables, bringing it into an inconsistent state where it performs unwanted actions (\eg, double-sending tokens to the adversary) that would not be possible in consistent states. 
Reentrancy attacks can be countered by using 
\href{https://docs.soliditylang.org/en/latest/security-considerations.html#reentrancy}{design patterns} ensuring that state updates are applied before potential reentrant calls, or by making contract calls mutually exclusive. 
However, systematically taking care of every call in a contract (including the pure transfers of currency)
is quite demanding and error-prone.

Reentrancy attacks are dealt with in various ways by the other platforms considered in this survey.
In \textbf{Solana}, reentrancy attacks are still possible but limited by the fact that re-entry is possible only as self-recursion.
In \textbf{Cardano}, reentrancy is ruled out by the absence of contract calls. 
The same goes with \textbf{Aptos}: invoking another contract is not possible unless its module is known at compile-time, and mutual recursive calls between modules are forbidden at compile-time.
Combined with the absence of callbacks or delegate calls, this rules out reentrancy by design.
\textbf{Algorand} is not vulnerable to reentrancy attacks, 
because, even though contract-to-contract calls are possible, a contract cannot call itself, even indirectly.
%
In \textbf{Tezos}, as already mentioned in~\Cref{sec:comparison:programming-style}, reentrancy attacks are mitigated by the fact that the caller function must complete, committing to its state, before performing other calls. 

Besides reentrancy, different smart contract languages/platforms suffer from different security concerns.
\textbf{Solana}, in particular, is prone to weaknesses related to its stateless model, which requires contract callers to provide the account containing the data to be read/written by the contract. 
Omitting some proper validations on accounts passed as input is a source of attacks: a notable example was the \href{https://arstechnica.com/information-technology/2022/02/how-323-million-in-crypto-was-stolen-from-a-blockchain-bridge-called-wormhole/}{wormhole attack},
which caused a loss of more than \$320 million~\cite{SmolkaGWDDKP23ccs}.
%
A specific vulnerability of this kind is the \emph{absence of signer verification}. 
Besides checking that the provided account is valid for a specific operation,
the contract must ensure that the transaction is signed by the \textit{holder} of that account. 
Omitting this check can lead to vulnerabilities.
For instance, if the developer omits this check in the $\txcode{win}$ method of the \contract{Bet} contract in~\Cref{fig:bet:acct-stateless},
then a malicious player could provide the oracle address without the corresponding signature, and set itself as the winner (bypassing the oracle altogether).
%
A related weakness is the \emph{absence of ownership verification}. 
For example, assume that the ownership check is omitted in the $\txcode{timeout}$ method of the stateless \contract{Bet} contract in \Cref{fig:bet:acct-stateless}. Then, an adversary could call $\txcode{timeout}$ with a specially-crafted  account that allows him to withdraw the whole pot.
By confirming that only the contract itself can modify the stored information, the data integrity remains protected.
Finally, the ability to invoke malicious or counterfeit contracts inside another contract invocation stems from the user's capability to supply any contract account, prompting the need for measures to verify the authenticity of the invoked contracts.


\textbf{Aiken} follows the approval style, in that the contract must check  the transaction fields to decide whether to approve an incoming transaction or not.
Forgetting even a single check may give rise to security vulnerabilities, possibly allowing an adversary to set a data field of the new state to an arbitrary value.
The same concerns apply to \textbf{PyTeal}, when used to write (approval-style) smart signatures.


Most of \textbf{Algorand} weaknesses revolve around its peculiar treatment of memory. In order to disincentivise the abuse of on-chain storage, every account must maintain a minimum balance that varies depending on how much memory it is using in the blockchain (which, in turn, depends on the number of distinct assets owned, contract data stored, \etc). Managing this balance constraint is tricky: developers must make sure that accounts the contract interacts with (and the contract account itself) always satisfy the minimum balance. This can create problems as transactions may unexpectedly fail, as they may lead the contract (or another account) to hold a balance lower than the allowed minimum. In particular, when emptying a contract account, it is essential to distinguish the case in which assets are sent from the case in which the contract account is closed. 

Further security implications of the fee mechanism design are discussed later in~\Cref{sec:comparison:fees}.


\subsection{Compile-time checks}
\label{sec:comparision:types}

With the exception of PyTeal/Algorand, all the languages considered in this paper feature strong typing.
\textbf{Solidity} supports subtype polymorphism, allowing programmers to implement contracts by inheriting other contracts in an object-oriented fashion.
It also features static visibility modifiers for functions and state variables, and dynamic (programmable) modifiers to restrict access to functions depending on run-time parameters.
Extra static checks performed by the compiler detect potential overflows/underflows and division by zero, 
\href{https://docs.soliditylang.org/en/latest/security-considerations.html#call-stack-depth}{stack size limit} vulnerabilities, and unwanted \href{https://swcregistry.io/docs/SWC-119/#shadowinginfunctionssol}{variable shadowing} caused by inheritance.
%
\textbf{Rust}
supports object-orientation, subtyping and parametric polymorphism.
Its compiler also tracks references and data ownership, ensuring memory safety and preventing data races.  
While Rust is a safe language, Solana does not provide an interface of the same quality, imposing several weakly typed programming patterns for writing contracts.
This renders the powerful checks performed by the Rust compiler irrelevant to some extent.
\textbf{Move}'s resource-oriented programming model is inspired by Rust: 
ownership of data is explicitly defined and enforced by the type system, and a borrow checker similar to Rust's prevents multiple mutable references to the same resource.
Linear types further add to the number of static checks by preventing code from replicating or losing currency and assets in general, ultimately mitigating double spending through typing.
Such features are similar to Rust's in principle, though in Move they are more integrated with the language syntax and straightforward for the programmer.
That is actually due to the fact that Move is a special-purpose language specifically tailored for asset management in smart contract programming.
\textbf{Aiken} too is a special-purpose language that stands out of the pack, as it delivers a purely functional style, with static typing and type inference.
Although this is fundamental to the safety of the validator script, static typing alone is not sufficient to rule out logic errors, as discussed in~\Cref{sec:comparison:security}.

\textbf{SmartPy} and \textbf{PyTeal}, although both based on Python, are substantially different when it comes to static checks.
PyTeal contracts are just Python programs that produce TEAL bytecode when executed. Instead, SmartPy contracts are compiled into Michelson, a typed bytecode language. The static typing and type inference supported by SmartPy are preserved by the compilation through type reconstruction. 
Furthermore, SmartPy contracts can carry type annotations, accessible as structured values through an API.
These are actually runtime entities for Python but are converted into type annotations in Michelson at translation time.
Such a hybrid approach improves the safety of SmartPy while retaining the simplicity of the Python syntax.
At the time of writing, Algorand lacks a compelling high-level language with static typing.
Programming in TEAL is equivalent to coding in assembly, thus with little to no static checks on the code. 
Although PyTeal features a rudimentary type system, type errors are still possible when encoding or decoding stored data, possibly leading to unpredictable errors and mishandling of the required datatypes.

Overall, with the notable exception of Move linear types, which can prevent double-spending, the type systems of the other languages can mostly prevent bad coding practices rather than some forms of vulnerability.
As noted in~\Cref{sec:comparison:security}, language design, when specifically tailored to rule out certain kinds of attacks in the first place, is more effective than most common forms of typing.



\subsection{Contract analysis and verification}
\label{sec:comparision:verification}

While compile-time checks are useful to rule out vulnerabilities due to common programming errors, they cannot guarantee that a contract respects some ideal behaviour in the presence of adversaries. 
Several tools have been developed to detect potential vulnerabilities in contracts. 
%
This is especially true for \textbf{Ethereum}, where dozens of bug detection tools with varying detection capabilities exist~\cite{Kushwaha22access,Ivanov23csur,GarfattaKGG21acsw}.
%
In \textbf{Solana}, current security tools include VRust~\cite{CuiZGT022ccs} and FuzzDelSol~\cite{SmolkaGWDDKP23ccs}. Both tools can detect Solana-specific vulnerabilities, like \eg the absence of signer checks and owner checks discussed in~\Cref{sec:comparison:security}, using different techniques
(inter-procedural data flow analysis for VRust, 
coverage-guided fuzzing for FuzzDelSol).
%
%
In \textbf{Algorand}, current tooling includes Panda~\cite{SunLZ23uss}, which is based on symbolic execution of TEAL code, and~\href{https://github.com/crytic/tealer}{Tealer}, which searches suspicious patterns in the control-flow graph extracted from the TEAL code.

More sophisticated tools enable the verification of contract implementations against an ideal, abstract description of their behaviour. 
For \textbf{Solidity}, this kind of analysis is partially supported by the assertion checker incorporated in the compiler, and by a few external analysis tools~\cite{BFMPS24fmbc,Wesley22vmcai}.
However, due to the intricacies of the Solidity/EVM semantics, these tools have several limitations in their precision and expressiveness of target properties~\cite{BFMPS24fmbc}.
\textbf{Move} features a property specification language that can be used by programmers to annotate function invariants.
Such invariants are verified at compile time through by the Move Prover, which is bundled with the Aptos toolchain.
A bytecode verifier validates compiled contracts at deploy-time, preventing maliciously crafted code from being uploaded to the blockchain.
Notably, the bytecode verifier enforces the same type-safety properties (including linearity) that are enforced by the Move compiler over the original source code. 
The work~\cite{Park24fmbc} applies the Move Prover to the formal verification of relevant functional requirements of modules of the Aptos Framework.   

Verification tools for \textbf{Tezos} include MiChoCoq and ConCert, which verify the functional correctness of contracts against a  specification based on pre- and post-conditions in the Coq proof assistant~\cite{MiChoCoq20,ConCert22}.
Other static analyzers exist, based on refinement types~\cite{Helmholtz22} and on abstract interpretation~\cite{OlivieriNAJS24dlt,BauMBB22}.

\subsection{On-chain / off-chain interactions}
\label{sec:comparision:onchain-offchain}

Off-chain systems 
are essential to extend blockchain features (\eg layer 2 and blockchain interoperability protocols) and provide users with Web3 services and decentralized applications.
Blockchain features (both at the consensus and at the contract layers),  contract languages, and off-chain libraries all impact the development of off-chain systems.
In particular, the interaction between off-chain and on-chain systems depends on how data flows to/from contracts.
In account-based platforms, data can be fed to contracts through method invocations. Some systems (Rust/Solana and Algorand) only allow base types as parameters in calls to contract entry-point functions, whereas Solidity, Anchor/Solana and Beaker/Algorand also support structured data. 
Outputting data from contracts is usually done through return values of method calls.
In the platforms where return values are not supported, contract outputs can be either written in other accounts (Solana) or embedded in transaction data (\eg, Algorand, Tezos, Cardano).  

Depending on where contract outputs are written, off-chain components use different techniques to retrieve them. 
In Move and SmartPy, the off-chain component can directly call methods because the blockchain preserves the interface and types. In the other systems, the off-chain components must first code the contract public interface before calling a method.
%
The other output retrieval technique is based on listening to events (or logs) emitted by the contract.
This is fully supported in Solidity, Move, and SmartPy. 
Contracts in PyTeal do not emit events but, as mentioned, rely mainly on the log to output results.
The other languages considered in this survey do not support events emissions. In Anchor and Aiken, low-level transaction logs can be exploited instead. 

Programming off-chain systems is facilitated by official or third-party supported SDKs and libraries (namely Web3). 
Solidity has stable support in a wide range of programming languages (including those for embedded devices). 
There are different versions of web3-like libraries and SDKs available for all of the other platforms examined. In particular, JavaScript libraries exist for Rust, Aiken (in two independent projects: Lucid and Mesh), Move, and SmartPy.
Python libraries provide support for Aiken, PyTeal, Move and SmartPy (Taquito). Rust libraries are available for Rust and Algorand.


\subsection{Fees}
\label{sec:comparison:fees}

The fee model established by the contract layer has non-negligible repercussions on the programming of smart contracts: developers must have a good understanding of the fee model in order to avoid paying more fees than strictly needed or incurring in potentially insecure programming patterns.

In \textbf{Ethereum}, fees depend on the sequence of EVM instructions needed for executing a transaction, and are paid by its sender.
Each EVM instruction has a cost, specified in terms of \emph{gas units}. 
The fee is the total amount of gas units consumed 
to execute the transaction times the price for gas unit.
The number of gas units per transaction is bounded:
transactions exceeding such limit pay the fee,
but have no other effects on the blockchain state.
So, although contracts can have unbounded loops and recursion,
in practice all computations are bounded.
The gas limit also bounds the contracts size, making it necessary to \href{https://ethereum.org/en/developers/tutorials/downsizing-contracts-to-fight-the-contract-size-limit/}{downsize the contract code} or distribute its logic across multiple contracts. 
The gas mechanism is a notorious source of attacks.
At the network level, DoS attacks~\cite{UnderPricedDoS} exploit the discrepancy between the gas units associated to EVM instructions and the actual computational resources needed for their execution~\cite{Perez20ndss}. 
Dealing with these attacks caused several revisions of the gas costs (\eg, \href{https://eips.ethereum.org/EIPS/eip-150}{EIP150}, \href{https://eips.ethereum.org/EIPS/eip-1559}{EIP1559}, \href{https://eips.ethereum.org/EIPS/eip-2929}{EIP2929}), possibly breaking existing contracts that depend on gas costs.
Fees can also be the basis for attacks to contracts. 
\Eg, a contract with a method that iterates over a dynamic data structure, such as a key-value map, can be attacked by making the structure grow until the iteration exceeds the maximum gas limit: in this way, the contract gets stuck, and its funds frozen. 
By combining the fee mechanism with transaction-ordering dependence,  attacks based on the unpredictability of fees are possible: \eg, an adversary might front-run a transaction to change the contract state so to cause the transaction to be reverted or pay more fees than expected.
The gas mechanism adopted by \textbf{Tezos} is conceptually similar to that of Ethereum, and therefore suffers from similar issues.

%
These attacks are not possible in the platforms where fees are predictable, as in Solana, Cardano and Algorand.
In \textbf{Solana}, transaction fees are determined solely by the number of required signatures for a transaction, rather than the amount of resources used. 
Besides transaction fees, Solana imposes fees on the data stored in accounts, to incentivize users not to waste on-chain space. 
If an account has not enough balance to cover the rent, it faces removal. Accounts can be exempted from paying fees by holding a balance that is at least equivalent to two years' worth of rent.   
Taking rent fees into account influences contract development in Solana.  
\Eg, in the stateless \contract{Bet} contract (\Cref{fig:bet:acct-stateless}), upon the completion of a final action 
the developer should 
close the storage account $\txcode{s}$ and return the remaining value, used for rent, back to the initializer. This requires an explicit coding of additional operations into the contract.

In \textbf{Cardano}, transaction fees depend on the transaction size and on the number of CPU steps and memory needed for its execution. All these data are \href{https://docs.cardano.org/cardano-testnet/tools/plutus-fee-estimator}{predictable} before sending the transaction, since Cardano 
is not subject to transaction-ordering dependencies, being based on the UTXO model.

%
In \textbf{Algorand}, although contracts are executed after compilation to low-level code as in Ethereum and Tezos, transaction fees are calculated differently. Namely, while in Ethereum and Tezos the fees depend on the sequence of executed low-level instructions, in Algorand they are determined only by the transaction size, with a lower bound set by the platform.


The fee model in \textbf{Aptos} incorporates elements from the models proposed by EVM, Algorand, and Solana, featuring a base minimum fee along with computation costs (referred to as I/O costs) and ``storage rent'' fees. Aptos transactions require a two-component fee structure that includes execution I/O costs and storage fees. The computation costs are measured in gas units, with the price fluctuating based on the network's load. The storage component is priced at a fixed rate in the platform's principal cryptocurrency. The storage fee can be refunded when the allocated space is deleted (as in Solana).



\input{tab-features.tex}

\subsection{Native \emph{vs.} programmable functionalities}
\label{sec:comparison:functionalities}

Developing our smart contracts benchmark was instrumental in understanding how different patterns are rendered in different languages/platforms. 
We detail in our repository~\cite{smart-contracts-comparison} the workarounds we adopted to implement the use cases in the various languages, and summarize below our main findings. 
For a quick reference, \Cref{tab:functionalities-vs-workarounds} depicts a comprehensive recap of the functionalities discussed below, plus a number of additional minor ones, for each language/platform explored in this paper.

\paragraph{Custom tokens}

All blockchain platforms come with a principal cryptocurrency (\eg, ETH for Ethereum), which is minted under the control of the consensus protocol and is exchangeable among users via direct transfers or programmatically via smart contracts. 
Many real-world contracts use tokens to represent custom assets 
(in our benchmark, the \href{https://github.com/blockchain-unica/smart-contracts-comparison/tree/main/contracts/token_transfer}{Token transfer} use case).
%
Unlike the principal currency, the minting of these custom tokens is not regulated by the consensus protocol, but rather by a user-defined policy. 
%
%
Solana, Cardano, Algorand, Tezos and Aptos support tokens natively, and allow contracts to define their transfers similarly to the native cryptocurrency.
In Move, which supports parametric polymorphism, custom assets are implemented via the generic type \mbox{\lstinline{Coin}}, whose  type parameter \lstinline{CoinType} specifies the fungible asset type.
Programmers can ensure that only assets of the same type are exchanged: this is achieved through the combined action of static typing and a dynamic lookup mechanism of resources driven by types.
In Ethereum, instead, tokens are not supported natively, and so they must be programmed as contracts,
by implementing standard interfaces
(\eg ERC20/ERC721 for fungible/non-fungible assets).
%
This comes at a cost for developers, as it is their duty to prevent asset duplication, unintended losses and other mishandling.
Furthermore, malicious token implementations could be an attacks vector to smart contracts~\cite{Tokenscope19ccs}.



\paragraph{Multisig transactions}

Another discriminating feature is given by \emph{multisig} transactions, \ie transactions that can carry the signature of multiple users. They are required \eg in the \href{https://github.com/blockchain-unica/smart-contracts-comparison/tree/main/contracts/bet}{Bet} use case, where two players must simultaneously deposit 1 token to join the game. 
The Cardano and Solana implementations fully respect the specification, since the underlying platforms support multisig transactions. 
\chreplaced
{Platforms such as Tezos and Ethereum do not support this feature natively but a workaround exists.
The implementations of \href{https://github.com/blockchain-unica/smart-contracts-comparison/tree/main/contracts/bet}{Bet} have to diverge from the specification by implementing a specific pattern that splits the join action into two steps.
}
{The other platforms do not support this feature, and so their implementations of \href{https://github.com/blockchain-unica/smart-contracts-comparison/tree/main/contracts/bet}{Bet} had to diverge from the specification, by splitting the join action in two steps.}
An alternative workaround is to implement a \emph{multisig contract}, which performs some given actions only if authorized by at least a given number of users.
In Tezos, \href{https://gitlab.com/tezos/tezos/-/blob/master/michelson_test_scripts/mini_scenarios/generic_multisig.tz}{multisig contracts} can be crafted by exploiting lambdas and the ability to verify signatures on arbitrary messages 
(furthermore, they are \href{https://tezos.gitlab.io/user/multisig.html}{natively supported} by the official client).
\chadded{Algorand supports multisig through multi-signature accounts, that is special sender addresses that have to be created by off-chain code and require a set of signatures to be authorized to perform the transaction. 
Aptos offers a similar mechanism based on off-chain code.
}

\paragraph{Contract updates}

Among the platforms considered in this paper, only Solana and Algorand allow to update contracts once deployed.
In the other platforms, this feature can be simulated through an
\href{https://github.com/blockchain-unica/smart-contracts-comparison/tree/main/contracts/upgradeableProxy}{UpgradeableProxy} contract, which intermediates the interactions between callers and a callee, allowing the owner to update the callee address (and so, the contract that processes function calls). 
The Solidity implementation exploits \href{https://docs.soliditylang.org/en/v0.8.16/introduction-to-smart-contracts.html#delegatecall-callcode-and-libraries}{delegate calls} to ensure that the caller and callee interact as there were no proxy intermediation.
In Solana, although contract updates are supported (at the cost of transferring account ownership to the new contract to remedy mutating restrictions), implementing the proxy does not seem possible. 
Cardano does not support contract calls, therefore the proxy cannot be implemented.
Still, contract updates are possible by making the validating script accept any transaction signed by the owner, allowing them to effectively replace the old script with the new one.
Aptos does not support contract calls or contract updates either.
%
SmartPy allows for contracts updates through lambda functions. Namely, the contract is represented as a mapping, whose values are lambdas. A method call is then translated into calling the corresponding lambda, while updating the contract is performed by updating the mapping.


\paragraph{Transaction batches}

In some use cases (\eg, circular trades and group payments) it is useful to batch transactions to ensure that either all or none of the transactions in a batch are executed.
Among the platforms considered here, transaction batching is supported natively only by Algorand (both client-side and contract-side) and by Solana (only client-side). 
In the absence of native support, a similar effect can be obtained by deploying a contract with a function that performs a \emph{specific} sequence of function calls. The \href{https://github.com/blockchain-unica/smart-contracts-comparison/tree/main/contracts/atomic_transactions}{atomic transactions} use case in our benchmark generalizes this by using a single contract that can process \emph{arbitrary} transaction batches. Our Solidity implementation exploits delegate calls to ensure transparency of the caller contract.
In Cardano, batching is not rendered in the strict sense of the term, but it is implicitly implemented by the UTXO mechanism. 
For instance, if we want two payments, say $1:\tokT$ from $\pmvA$ to $\pmvB$ and $1:\tokTi$ from $\pmvB$ to $\pmvC$, to happen atomically, we can obtain the same effect by a \emph{single} transaction with two inputs (one redeeming $1:\tokT$ with $\pmvA$'s signature and the other redeeming $1:\tokTi$ with $\pmvB$'s) and two outputs, controlling  $1:\tokT$ and $1:\tokTi$ with $\pmvB$'s and $\pmvC$'s signatures, respectively. 
Although Aptos does not support transaction batches natively, programmers can pack multiple actions in a single Move script, \ie a code block that can invoke functions defined in contract modules and is executed atomically (similarly to Ethereum's workaround). 


\paragraph{Time constraints}

Most platforms allow contracts to set time constraints by making the current block number or transaction timestamp readable by the contract. This is a common feature in real-world scenarios: in our benchmark, it occurs \eg in the
\href{https://github.com/blockchain-unica/smart-contracts-comparison/tree/main/contracts/bet}{Bet},
\href{https://github.com/blockchain-unica/smart-contracts-comparison/tree/main/contracts/auction}{Auction},  \href{https://github.com/blockchain-unica/smart-contracts-comparison/tree/main/contracts/crowdfund}{Crowdfund}, 
\href{https://github.com/blockchain-unica/smart-contracts-comparison/tree/main/contracts/htlc}{HTLC}, 
\href{https://github.com/blockchain-unica/smart-contracts-comparison/tree/main/contracts/vault}{Vault} and 
\href{https://github.com/blockchain-unica/smart-contracts-comparison/tree/main/contracts/vesting}{Vesting} use cases.
%
In Cardano, contracts cannot access the global blockchain state (including the block number), but time constraints can be implemented leveraging the validity interval field of transactions.
Knowing only a time interval rather than the exact time might introduce some approximations \wrt the ideal behaviour. 
\Eg, in the Aiken version of the \href{https://github.com/blockchain-unica/smart-contracts-comparison/tree/main/contracts/vesting}{Vesting} contract, the beneficiary can receive slightly less than the amount prescribed by the vesting function at the exact time the transaction is processed. This discrepancy arises since the amount is determined as a function of the (lower bound of) the validity interval.

\paragraph{Key-value maps, dynamic arrays, and bounded loops}

All the languages considered in this paper support dynamic data structures such as 
key-value maps and arrays.
In the stateful account-based models, they are stored in the contract account. 
In Solana, instead, values are stored in accounts, whose addresses serve as keys for key-value maps. For this specific purpose, Solana uses special addresses 
that are deterministically generated but lack corresponding private keys and are tailored to be under the exclusive control of a designated smart contract.
In Aiken/Cardano, dealing with dynamic data structures raises some efficiency concerns, since updating the contract state requires sending a transaction carrying the \emph{whole} new state. 
In contracts whose state involves arrays or key-value maps 
that can grow significantly during execution,
these transactions may become larger and larger.
This has two drawbacks, in that the transaction fees 
increase with the transaction size,
and in that there is a hard limit (16KB) on this size.
These issues could be potentially mitigated by using
cryptographic techniques (\eg, Merkle trees) to minimize
the amount of data stored on-chain.
Currently, this has to be implemented manually by the programmer,
as Aiken does not feature constructs to 
automatize the management of large data structures
exploiting these cryptographic primitives.
%
In Algorand, dynamic data structures can be implemented by using \emph{boxes}, \ie pieces of memory that can be allocated at any time during the lifetime of the contract, at the cost of an increased minimum balance for the contract account. These boxes are, however, fairly expensive. When a use case requires a key-value map indexed on account addresses, the use of the local storage of these accounts is preferred to that of boxes. Consider, for instance, a contract that receives deposits from users, and that needs to record the amounts transferred by each user (like \eg in the \href{https://github.com/blockchain-unica/smart-contracts-comparison/tree/main/contracts/crowdfund}{crowdfund} use case in our benchmark). In Algorand this can be achieved by distributing the map across the local storage of each account depositing tokens to the contract.
As a single call can only read the content of a limited number of boxes (8 per call), it is not possible to iterate over structures that span a large number of boxes. This means that iterating over arrays is still feasible provided that the array is encoded in a single box; instead, iterating over dynamic data structures such as key-value maps is quite problematic. Furthermore, working with multiple maps is tricky: since box storage maintains a single key-value store, making it appear as multiple maps requires the developer to manually handle the partitions. This issue, together with the varying minimum balance on the insertion of new key-value pairs, makes the use of key-value maps in Algorand quite burdensome.

\paragraph{Randomness}

Some use cases require randomly-generated values (\eg, in lotteries and other games to choose a winner or to draw a card). 
\chadded{Randomness has also proven effective in mitigating the threats posed by criminal smart contracts~\cite{WangBLLCZ19isci}.}
Although some centralized randomness beacons are available, their use is not considered secure, since dishonest providers can bias their outputs~\chadded{\cite{BlautMW23icbc2,QianHLWLWZH23tse}}. A secure alternative is given by \emph{commit-reveal-punish} schemes, which construct random values by combining values independently provided by users. To ensure that no one can observe the others' values to craft their own (which would easily lead to attacks), these schemes force users to commit the hashes of the chosen values before revealing them and use collaterals to rule out dishonest users who do not reveal (see \eg the \href{https://github.com/blockchain-unica/smart-contracts-comparison/tree/main/contracts/htlc}{HTLC} use case). A drawback of these schemes is that they become quite complex when many users are involved. A viable alternative is given by Verifiable Random Functions~\cite{Micali99focs}, a cryptographic primitive that allows users to generate publicly verifiable random values. Among the platforms considered here, only Algorand offers this feature natively, by combining a randomness seed beacon and a special opcode to verify the correct generation of values.
\chadded{We note that the analysis and formal verification of randomized smart contracts is currently an under-explored research field, with limited tool support~\cite{Mazurek21fc}.}

%% file: tab-overview.tex
\begin{table*}[t]
\caption{Features of the contract layer of some of the main smart contract platforms.} 
\label{tab:overview}
\small
\centering
\begin{tabular}{cccccc}
\textbf{Platform} & \textbf{Accounting model} & \textbf{Contract storage} & \textbf{Fees depend on \ldots} & \textbf{Main contract languages} & \textbf{Programming style}
\\
\hline
Ethereum & Account-based & Stateful & Tx computation & Solidity & Procedural \\
Solana & Account-based & Stateless & Num. signers, Data size & Rust & Procedural \\
Cardano & UTXO & Stateful & Tx size, Tx computation & Plutus, Aiken & Approval \\
Aptos & Account-based & Stateless & Tx computation, Data size & Move & Procedural \\
Algorand & Account-based & Stateful & Constant & PyTeal & Procedural \\
Tezos & Account-based & Stateful & Tx computation, Data size  & SmartPy, Ligo & Procedural \\
\hline
\end{tabular}    
\end{table*}

%% file: tab-usecases.tex
\newcommand{\y}{\checkmark}
\newcommand{\Y}[1]{\checkmark${}^{#1}$}
\newcommand{\X}[1]{\workaround${}^{\!#1}$}

\newcolumntype{R}[2]{%
    >{\adjustbox{angle=#1,lap=\width-(#2)}\bgroup}%
    l%
    <{\egroup}%
}
\newcommand*\rot{\multicolumn{1}{R{45}{1em}}}


\begin{table*}[t]
\caption{\chadded{Functionalities required by the use cases in the benchmark. Entries marked with \workaround${}^i$ denote functionalities that can be used to implement workarounds in case the functionality marked with \checkmark${}^i$ is not natively available in the given language/platform.}}
\label{tab:functionalities-vs-usecases}
\small
{
 \resizebox{0.95\textwidth}{!}{
\begin{tabular}{p{2.8cm}ccccccccccccccccccc}
\textbf{Use case}
& \rot{Native tokens}
& \rot{Custom tokens}
& \rot{Multisig transactions}
& \rot{Contract updates}
& \rot{Transaction batches}
& \rot{Time constraints} 
& \rot{Key-value maps}
& \rot{Dynamic arrays}
& \rot{Bounded loops}
& \rot{Randomness}
& \rot{Transaction revert} 
& \rot{Contract-to-contract calls}
& \rot{Delegate contract calls}
& \rot{In-contract deployment}
& \rot{Hash on arbitrary messages}
& \rot{Versig on arbitrary messages}
& \rot{Bitstring operations}
& \rot{Arbitrary-precision arith.}
& \rot{Rational arith.}
\\
\hline
Bet                   & \y &    & \y &    &    & \y &    &    &    &    & \y &    &    &    &    &   &   &   & \\
Simple transfer       & \y &    &    &    &    &    &    &    &    &    & \y &    &    &    &    &   &   &   & \\
Token transfer        &    & \Y{1} &    &    &    &    & \X{1}  &    &    &    & \y & \X{1} &    &    &    &   &   &   & \\
HTLC                  & \y &    &    &    &    & \y &    &    &    &    & \y &    &    &    & \y &   &   &   & \\
Escrow                & \y &    &    &    &    &    &    &    &    &    & \y &    &    &    &    &   &   &   & \\
Auction               & \y &    &    &    &    & \y & \y &    &    &    & \y &    &    &    &    &   &   &   & \\
Crowdfund             & \y &    &    &    &    & \y & \y &    &    &    & \y &    &    &    &    &   &   &   & \\
Vault                 & \y &    &    &    &    & \y &    &    &    &    & \y &    &    &    &    &   &   &   & \\
Vesting               & \y &    &    &    &    & \y &    &    &    &    & \y &    &    &    &    &   &   &   & \\
Storage               &    &    &    &    &    &    &    & \y &    &    &    &    &    &    &    &   &   &   & \\
Simple wallet         & \y &    &    &    &    &    &    &    &    &    & \y &    &    &    &    &   &   &   & \\
Price bet             & \y &    &    &    &    &    &    &    &    &    & \y & \y &    &    &    &   &   &   & \\
Payment splitter      & \y &    &    &    &    &    & \y & \y &    &    & \y &    &    &    &    &   &   &   & \\
Lottery               & \y &    & \y &    &    & \y &    &    &    & \Y{1} & \y &    &    &    & \X{1} &   & \X{1} &   & \\
Const.~prod.~AMM      &    & \Y{1} &    &    &    &    & \X{1} &    &    &    & \y & \X{1} &    &    &    &   &   & \y & \y \\
Upgradeable proxy     &    &    &    & \Y{1} &    &    &    &    &    &    & \y & \X{1} & \X{1} &    &    &   &   &   & \\
Factory               &    &    &    &    &    &    & \y & \y &    &    & \y &    &    & \y &    &   &   &   & \\
Decentralized identity&    &    &    &    &    &    &    &    &    &    & \y &    &    &    &    \y & \y &   &   & \\
Editable NFT          &    & \Y{1} &    &    &    & \X{1} &    &    &    &    & \y & \X{1} &    &    &    &   &   &   & \\
Anonymous data        &    &    &    &    &    &    &    & \y & \y &    & \y &    &    &    &    &   &   &   & \\
Atomic transactions   &    &    &    &    & \Y{1} &    &    & \X{1} & \X{1} &    & \X{1} &    &    &    &    \X{1} & \X{1} &   &   & \\
\hline
\end{tabular}
}}
\end{table*}

%% file: tab-loc.tex
\renewcommand{\tabrotdegrees}{90}
\renewcommand{\tabrothspace}{0pt}

\begin{table}[t]
\small
\centering
\caption{Lines of code (LoC)\chadded{, excluding comments and empty lines,} of a selection of use cases implementations. For Solana we show  LoC of Anchor code, since it is more succinct than pure Rust.}
\label{tab:loc-comparison}
\begin{tabular}{p{2cm}ccccccc}
\textbf{Use case}       & 
\rotatebox{\tabrotdegrees}{\textbf{Solidity}}\hspace{\tabrothspace}
\rotatebox{\tabrotdegrees}{Ethereum}
& 
\rotatebox{\tabrotdegrees}{\textbf{Anchor}}\hspace{\tabrothspace}
\rotatebox{\tabrotdegrees}{Solana}
&
\rotatebox{\tabrotdegrees}{\textbf{Aiken}}\hspace{\tabrothspace}
\rotatebox{\tabrotdegrees}{Cardano}
& 
\rotatebox{\tabrotdegrees}{\textbf{PyTeal}}\hspace{\tabrothspace}
\rotatebox{\tabrotdegrees}{Algorand}
& 
\rotatebox{\tabrotdegrees}{\textbf{Move}}\hspace{\tabrothspace} 
\rotatebox{\tabrotdegrees}{Aptos}
& 
\rotatebox{\tabrotdegrees}{\textbf{SmartPy}}\hspace{\tabrothspace} 
\rotatebox{\tabrotdegrees}{Tezos}
\\
\hline
Bet             & 39    & 137   & 158   & 110   & 62    & 53            \\
Simple transfer & 18    & 91    & 120   & 49    & 30    & 21            \\
HTLC            & 25    & 123   & 115   & 60    & 49    & 31            \\
Escrow          & 41    & 176   & 120   & 94    & 45    & 28            \\
Auction         & 51    & 152   & 221   & 129   & 40    & 45            \\
Crowdfund       & 31    & 182   & 129   & 103   & 49    & 33            \\
Vault           & 40    & 166   & 171   & 103   & 57    & 38            \\
Vesting         & 39    & 149   & 125   & 105   & 48    & 28            \\
Storage         & 11    & 82    & 75    & 32    & 23    & 18            \\
Simple wallet   & 47    & 183   & 169   & 87    & 108   & 47            \\
\hline
\emph{Average} & \emph{34}	& \emph{144}  &	\emph{140}	& \emph{88} &	\emph{51}	& \emph{34}
\\
\hline
\end{tabular}
\end{table}

%% file: tab-features.tex
\begin{table*}
\centering
\caption{Functionalities supported by smart contract languages/platforms. The first group refers to the functionalities described in~\Cref{sec:comparison:functionalities}.
Checkmarks $\native$ denote functionalities that are available natively in the language or via the blockchain APIs. The symbol \workaround denotes functionalities that can be implemented in a smart contract with some practical workaround (\eg, ERC20/ERC721 interfaces for custom tokens in Ethereum). \chadded{An empty cell indicates that the functionality is not supported by the platform.} Further workarounds are implemented in our benchmark~\cite{smart-contracts-comparison}.}
\label{tab:functionalities-vs-workarounds}
\small
\begin{tabular}{p{5cm}cccccc}

\textbf{Functionalities}       & 
\rotatebox{90}{\textbf{Solidity}}
\rotatebox{90}{Ethereum} 
& 
\rotatebox{90}{\textbf{Rust/Anchor}}
\rotatebox{90}{Solana}
& 
\rotatebox{90}{\textbf{Aiken}} 
\rotatebox{90}{Cardano}
& 
\rotatebox{90}{\textbf{TEAL/PyTeal}} 
\rotatebox{90}{Algorand}
& 
\rotatebox{90}{\textbf{Move}}
\rotatebox{90}{Aptos}
& 
\rotatebox{90}{\textbf{SmartPy}}
\rotatebox{90}{Tezos}
\\
\hline

Native tokens  
& \native 
& \native 
& \native 
& \native  
& \native 
& \native      
\\
Custom tokens  
& \workaround 
& \native 
& \native 
& \native  
& \native 
& \native      
\\
Multisig transactions
& \workaround
& \native 
& \native 
& \native
& \native 
& \workaround  

\\
Contract updates
& \workaround 
& \native
&
& \native
& 
& \workaround 

\\
Transaction batches
& \workaround
& \workaround
& \native 
& \native  
& \workaround 
& \workaround

\\
Time constraints
& \native
& \native
& \native
& \native
& \native
& \native      

\\

Key-value maps \& Dynamic arrays
& \native
& \native
& \native 
& \native 
& \native
& \native    

\\

Bounded loops
& \native
& \native
& \native 
& \workaround
& \native
& \native    

\\

Randomness
& \workaround
&  \workaround
& \workaround
& \native
& \native
& \workaround

\\

\hline 

Transaction revert
& \native
& \native 
& \native 
& \native
& \native
& \native           

\\

Contract-to-contract calls 
& \native
& \native
& 
& \native
&  
& \native  

\\
In-contract deployment
& \native
&  
&
& \native 
& \native 
& \native     

\\

Delegate contract calls
& \native
&
&
&   
& \native 
&         

\\
Hash on arbitrary messages       
& \native
& \native 
& \native 
& \native 
& \workaround 
& \native    

\\
Versig on arbitrary messages
& \native 
&  
& \native
& \native 
& \native
& \native     

\\
Bitstring operations 
& \native
& \native
&
& \native
& \native 
& \native        

\\
Arbitrary-precision arithmetic
&
& \native 
& \native
&  
&  
&        

\\
Rational arithmetic
& \workaround
& \workaround
& \native 
& \workaround  
& \workaround 
& \native    

\\
\hline
\end{tabular}
\end{table*}

%% file: tab-strengths.tex
\begin{table*}[tbhp!]
\centering
\caption{Strengths and weaknesses of smart contract languages.}
\label{tab:strengths}
\small
\begin{tabular}{p{1.8cm}p{7cm}p{7cm}}
\hline
{\bf Language} & {\bf Strengths} & {\bf Weaknesses} \\
\hline
\begin{tabular}{l} \textbf{Solidity} \\ Ethereum \end{tabular} &
\begin{tabular}{l} 
Familiar procedural programming style \\
Extensive documentation \\ 
Rich ecosystem of analysis tools third-party libraries
\end{tabular} 
& 
\begin{tabular}{l}
EVM design induces vulnerabilities (\eg reentrancy) \\
Low-level interferences with the semantics (\eg, fees) \\
Transaction-ordering dependencies
\end{tabular}
\\[1pt]
\begin{tabular}{l} \textbf{Rust/Anchor} \\ Solana \end{tabular} &
\begin{tabular}{l} 
Powerful strongly typed language \\
Good parallelization / transaction throughput \\
Rich ecosystem and community \\ 
\end{tabular} 
& 
\begin{tabular}{l} 
Rust ownership system not really used \\ 
Verbose programming model \\
Need to serialize/deserialize data manually \\
\end{tabular}
\\[15pt]   
\begin{tabular}{l} \textbf{Aiken} \\ Cardano \end{tabular} & 
\begin{tabular}{l} 
Strongly-typed functional paradigm \\
No transaction-ordering dependencies (UTXO) \\
Arbitrary-precision arithmetic \\
\end{tabular} 
& 
\begin{tabular}{l} 
Requires reasoning about the structure of transactions \\
No contract calls \\
Transactions must specify the \emph{whole} contract state \\
\end{tabular}
\\[1pt]
\begin{tabular}{l} \textbf{TEAL/PyTeal} \\ Algorand \end{tabular} &
\begin{tabular}{l}  
Rich set of functionalities (transaction batches, \\
Verifiable Random Functions, contract updates) \\
Predictable transaction fees \\
\end{tabular} 
& 
\begin{tabular}{l} 
Cumbersome handling of memory (local, global, boxes) \\
Cumbersome constraint on minimum accounts balance \\
Weaker typing guarantees \wrt other languages
\end{tabular}
\\[1pt]
\begin{tabular}{l} \textbf{Move} \\ Aptos \end{tabular} &
\begin{tabular}{l} 
Linear types prevent errors (\eg double spending) \\ 
Static prover enforces semantic properties \\
Extensive stdlib and framework \\
\end{tabular} 
& 
\begin{tabular}{l} 
Requires good understanding of linear types \\ 
Lack of community and third-party libraries \\
\end{tabular}
\\[15pt]
\begin{tabular}{l} \textbf{SmartPy} \\ Tezos \end{tabular} &
\begin{tabular}{l} 
Strong typing and type inference on top of Python \\ 
Queued method call to avoid reentrancy attacks \\ 
Allows Python meta-code \\
\end{tabular} 
& 
\begin{tabular}{l} 
A strongly typed Python is a little awkward \\ 
Lack of dedicated third-party libraries and tools \\  
\end{tabular}
\\
\hline
\end{tabular}
\end{table*}

%% file: related.tex
\section{Related work}
\label{sec:related}

In recent years, there has been a surge in advancements within decentralized, permissionless blockchain technologies, with a particular emphasis on smart contracts. However, despite this remarkable progress, there remains a substantial gap in our understanding of the fundamental principles and programming paradigms that underpin smart contracts.
While numerous studies have examined specific applications and challenges associated with smart contracts, there has been a glaring absence in the exploration of their  programming principles and languages.

Recent literature reviews, such as \cite{sharma2023review,zheng2020overview}, aimed to provide systematic overviews of technical challenges in smart contract development. Sharma \emph{et al.}~\cite{sharma2023review} delved into aspects such as consensus algorithms, permission policies, Turing completeness, and data models. They identified crucial challenges such as readability, code correctness, execution efficiency, privacy concerns, and gas exceptions. Similarly, Zheng \emph{et al.}~\cite{zheng2020overview} performed a comparative analysis of platforms and applications, examining essential aspects such as creation, readability, execution efficiency, transaction ordering, deployment, and privacy-preserving mechanisms. They also categorized applications and outlined common use cases across diverse domains.
In contrast to \cite{sharma2023review,zheng2020overview}, 
which primarily rely on comparing results from existing literature on smart contract languages, 
our comparisons is based on our practical experience developing a  benchmark of use cases, where we contrast different platforms/languages by implementing a range of smart contracts in each language. This benchmarking methodology enables us to perform a comprehensive comparative analysis, offering insights into programming styles, readability and usability, compile-time checks, on-chain/off-chain interactions, as well as security considerations across different platforms/languages.

The impact of smart contracts on industry has spurred a wealth of research, see, \eg, \cite{varela2021smart,dhaiouir2020systematic}. Varela-Vaca \emph{et al.}' work \cite{varela2021smart} categorised smart contract languages from both academic and industrial perspectives, with an emphasis on improving developer experiences for creating more human-readable smart contracts. 
Similarly, Dhaiouir \emph{et al.}' literature review \cite{dhaiouir2020systematic} compared distributed platforms, aiming to assist businesses in selecting suitable platforms for blockchain-based applications, thus facilitating informed decision-making.
Vacca \emph{et al.}~\cite{vacca2021systematic} reviewed methods, techniques, and tools for improving the design, construction, testing, maintenance, and overall quality of smart contracts and DApps. Similarly, Zou \emph{et al.}~\cite{zou2021smart} performed an empirical study on developers' challenges and practices in smart contract development, with a focus on Ethereum. They collected valuable insights into the current state of the art through interviews and surveys with industry practitioners.

A few works address smart contract languages for UTXO blockchains, mainly focussing on Bitcoin and Cardano. 
Outside academic research, Bitcoin is quite unattractive as a layer-1 smart contract platform, because of the expressiveness limitations of its script language, its low throughput and high transaction fees. 
Still, a small subset of the use cases in our benchmark can be implemented also on Bitcoin~\cite{AtzeiBCLZ18post}, either using Bitcoin script of higher-level languages such as BitML~\cite{AtzeiBLYZ19fse}.
In the Cardano literature, the work \cite{Brunjes20isola} draws an interesting comparison between the account-based and the UTXO model based on the implementation of a token use case in Solidity and in Plutus.
In this paper we extend the comparison in~\cite{Brunjes20isola}, by implementing a large set of use cases in six smart contract languages.
The relation between transaction redeem scripts and structured contracts in Cardano is explored in recent research~\cite{Vinogradova24fmbc}.

Several surveys address the challenges related to security vulnerabilities.
Hu \emph{et al.}~\cite{hu2021comprehensive}  categorized schemes and tools aimed at improving secure smart contract development.   Additionally, they addressed challenges like privacy breaches, execution inefficiencies, and contract complexity by categorizing extensions and alternative systems for contract execution.
Rouhani \emph{et al.}~\cite{rouhani2019security} conducted an extensive review focusing on smart contract platforms and domain-specific programming languages, focussing on security vulnerabilities and performance optimization. Their study explored methods and tools for mitigating vulnerabilities.
Hewa \emph{et al.}~\cite{hewa2021survey} undertook a comprehensive survey on smart contracts, emphasizing aspects like security, privacy, gas cost, and concurrency. In particular, they explored the integration of smart contracts with emerging technologies such as artificial intelligence and game theory.

The works \cite{voloderA23icbc,parizi2018smart}, which compare  smart contract languages in terms of usability and security, are the most closely aligned with ours. In~\cite{voloderA23icbc}, Voloder \emph{et al.}~conducted a comparative analysis of five platforms focusing on developers' perspectives. Their comparison examines critical features such as documentation availability, ease of installation, automated testing capabilities, implementation efforts, and the required level of expertise for specific use cases and chains.
Parizi \emph{et al.}~\cite{parizi2018smart} analysed the usability and security aspects of three smart contract languages: Solidity, the \href{https://docs.kadena.io/pact/}{Pact} language for Kadena (which is Turing-incomplete), and \href{https://liquidity-lang.org/}{Liquidity} for Tezos (a typed functional language).  
The paper offers a comparative analysis of these languages, demonstrating sample contract implementations and evaluating them in terms of usability and security. 
In contrast to \cite{parizi2018smart}, we opt to exclude Pact and Liquidity from our selection of smart contract language. This choice is based on our paper's emphasis on Turing-complete languages (which is also the reason why we neglect contract languages on Bitcoin), as well as the recognition that Liquidity is no longer actively maintained. 

Differing from prior research, this paper offers a unique perspective by providing a detailed technical comparison of smart contract languages from the standpoint of programming language theory. We delve into programming styles, language constructs, and typing considerations, complemented by a qualitative assessment derived from hands-on experience in crafting a standardized benchmark for smart contracts. Marking a pioneering effort, this work provides the first extensive hands-on evaluation, facilitating both comparison between smart contract languages and analysis of development and execution costs.


%% file: conclusions.tex
\section{Conclusions}
\label{sec:conclusions}

We have compared the smart contract languages of some of the most widespread blockchains. The comparison, which was performed along different axes, is based both on the literature and on hands-on knowledge derived from the construction of a common benchmark of smart contracts. \Cref{tab:strengths} summarises the main findings of our comparison: we conclude by discussing the lessons learned in our work.

\paragraph{Lesson learned \#1: language abstractions}

Our analysis highlights the need for high-level abstractions over the low-level details of the underlying blockchain. Clean abstractions are crucial to simplifying reasoning about the correctness and security of contracts. Not all languages considered in this paper have such clean abstractions. For instance, the lack of good abstractions for 
tokens and contract-to-contract interactions is one of the main causes of vulnerabilities in Solidity/Ethereum contracts (see~\Cref{sec:comparison:security}). The lack of good abstractions over the transactions level in Aiken/Cardano induces a burdensome programming style for contracts in these languages, with potentially harmful consequences on their security (see~\Cref{sec:comparison:programming-style,sec:comparison:security}). Furthermore, the interference between the low-level fee mechanisms and the contract semantics is not always hidden from programmers, who must have a good understanding of these mechanisms to avoid writing inefficient or vulnerable contracts (see~\Cref{sec:comparison:fees}).


\paragraph{Lesson learned \#2: typing assets}

Assets deserve special treatment at the type level in order to prevent programmers from making financial mishaps when manipulating crypto-assets.
This can be enforced to varying degrees.
The loosest form is to represent assets by means of a custom datatype (distinct from the plain integer type), which prevents programmers from performing unwanted arithmetic operations on assets.
By limiting the number of possible operations for the asset datatype, and providing only a minimal set of primitives for transferring assets, account-based platforms can reduce error-proneness when handling valuable tokens.
Disciplining assets and transactions in UTXO platforms is not as straightforward, though.
In Aiken, for instance, asset transfers are implemented as record field updates where arithmetic operations are required to manipulate amounts.
A special asset datatype with its own set of functions would make things harder and verbose for the programmer.
The strictest form of control among the languages reviewed in this paper is Move's linear types, which push the envelope by forbidding duplication and loss of assets at compile-time (see~\Cref{sec:comparison:programming-style,sec:comparision:types}).
Although such a strict type discipline is hard to digest for a casual programmer, from our experience it does not come without its own merits.
Move contracts seem less susceptible to asset-related issues (\eg double spending and financial loss) compared to other platforms, underlining that smart contract languages ought to dare more than general-purpose languages when it comes to the discipline imposed on types, especially on the type representing assets.

\paragraph{Lesson learned \#3: native vs.~programmable \mbox{functionalities}}
The smart contract languages considered are characterised by different sets of native functionalities, as displayed in \Cref{tab:functionalities-vs-workarounds}. The absence of some functionality could be detrimental to the implementation of certain use cases, making it either impossible, or possible only through complex workarounds and adaptations of the requirements. We have directly experienced the lack of native functionalities in our benchmark, where some implementations required such adaptations and workarounds.
Although in principle all the languages considered in this paper are Turing-powerful (up-to computation bounds due \eg to transaction fees), some work\-arounds could be extremely impractical due to the high costs of on-chain computation and storage, besides the computation bounds. For instance, implementing arbitrary-precision arithmetic via Church encodings would make little sense.
Improper workarounds could affect security and decentralization. This is the case, \eg, of generating randomness via block timestamps or external oracles. 
In general, the availability of specific native functionalities could be an important factor in the decision-making process to choose a blockchain platform, among others~\cite{Farshidi20tem}.

\paragraph{Lesson learned \#4: procedural \emph{vs.}~approval style}

As we have seen in~\Cref{sec:comparison:programming-style},
smart contract languages can be partitioned into two classes based  on the programming style they support:
the \emph{procedural style}, where contracts react to transactions by updating their state and triggering effects (\eg, token transfers), and the \emph{approval style}, where transactions already contain their effect, and the contract reacts by deciding whether to approve a transaction or not, depending on its state and by the environment.
In~\Cref{sec:comparison:loc,sec:comparison:security} we have seen that the programming style has deep implications on the readability of contracts and on their security: roughly, the approval style is less readable and more error-prone, since the programmer must ensure that the new state is a correct update of the old one, which might involve multiple checks on the transactions fields.
Based on the implementation of our benchmark, we argue that the procedural style is overall the most practical, even though in some of its incarnations we note that the produced code is burdened with boilerplate code (\eg, in Rust/Solana), or with type-based manipulations of resources that may look unfamiliar to average programmers (\eg, in Move/Aptos). 
An open question is whether it is possible to reconcile the procedural style with the UTXO-based model, so to program smart contracts \emph{\`a la} Solidity while preserving the key strengths of UTXO blockchains like Cardano (\eg, the absence of transaction-ordering dependencies and the parallelizability of transactions).